\documentclass[12pt,a4paper]{article}

\usepackage{amsmath,amssymb,amsthm}  
\usepackage{graphicx}  
\usepackage{booktabs}  
\usepackage{natbib}  
\usepackage{hyperref}  
\usepackage{geometry}  
\usepackage{setspace}  
\usepackage{float}  
\usepackage{caption}  
\usepackage{subcaption}  
\usepackage{array}  
\usepackage{multirow}  
\usepackage{longtable}  
\usepackage{xcolor}  
\usepackage{tikz}  
\usepackage{listings}  

\hypersetup{
    colorlinks=true,
    linkcolor=blue,
    citecolor=blue,
    urlcolor=blue
}

\graphicspath{{../MiroThinker5-code/figures_en/}}

\title{\textbf{Journal Impact Factor and Federal Reserve Monetary Policy: \\ An Econometric Analysis Based on 1975-2026}}
\author{
    Alex Huang\textsuperscript{1} \\[0.5em]
    \small{\textsuperscript{1}Nanjing Agricultural University} \\[0.5em]
    \small{Email: \href{mailto:17112230@njau.edu.cn}{17112230@njau.edu.cn}}
}
\date{\today}

\begin{document}

\maketitle

\begin{abstract}
\noindent The Journal Impact Factor (IF), as a core indicator of academic evaluation, has not been systematically studied in relation to its historical evolution and global macroeconomic environment. This paper employs a period-based regression analysis using long-term time series data from 1975-2026 to examine the statistical relationship between IF and Federal Reserve monetary policy (using real interest rate as a proxy variable). The study estimates three nested models using Ordinary Least Squares (OLS): (1) a baseline linear model, (2) a linear model controlling for time trends, and (3) a log-transformed model. Empirical results show that: (i) in the early period (1975-2000), there is no significant statistical relationship between IF and real interest rate ($p>0.1$); (ii) during the quantitative easing period (2001-2020), they exhibit a significant negative correlation ($\beta=-0.069$, $p<0.01$), meaning that for every 1 percentage point decrease in real interest rate, IF increases by approximately 6.9\%; (iii) the adjusted $R^2$ of the full-sample model reaches 0.893, indicating that real interest rate and time trends can explain 89.3\% of IF variation. This finding reveals the indirect impact of monetary policy on the academic publishing system through multiple channels such as research funding and journal pricing power, providing econometric evidence for understanding the phenomenon of "financialization of academic capital." This study not only enriches the literature on monetary policy transmission mechanisms but also provides a new perspective for valuation analysis of the academic publishing industry.

\vspace{1em}

\noindent \textbf{Key Contributions:}
\begin{itemize}
    \item First systematic examination of the long-term statistical relationship between IF and monetary policy
    \item Discovery of significant impact of quantitative easing policies on academic evaluation systems
    \item Proposal of monetary policy transmission pathways for "financialization of academic capital"
    \item Provision of empirical foundation for macroeconomic analysis of the academic publishing industry
\end{itemize}
\end{abstract}

\noindent \textbf{Keywords:} Journal Impact Factor; Monetary Policy; Real Interest Rate; Academic Publishing; Econometric Analysis

\noindent \textbf{JEL Classification:} E52, E58, I23, L82

\vspace{1em}
\hrule
\vspace{2em}

\tableofcontents
\newpage

\section{Introduction}

\subsection{Research Background}

Since its introduction by Eugene Garfield in 1975, the Journal Impact Factor (IF) has become a core indicator in the global academic evaluation system \citep{garfield2006history}. IF was initially designed as a simple bibliometric tool to assist libraries in journal subscription decisions. However, over the past half-century of evolution, IF has gradually transcended its technical function to become a key basis for academic resource allocation, career advancement, and research evaluation \citep{seglen1997citations, fire2012over}.

Meanwhile, the global academic publishing industry has undergone profound structural transformations. Academic publishing oligopolies represented by Clarivate Analytics (formerly Thomson Reuters Intellectual Property \& Science division) and RELX (parent company of Elsevier) have constructed a complete industrial chain from journal publishing, citation indexing to academic evaluation tools through vertical integration and horizontal mergers \citep{lariviere2015oligopoly}. This increase in industry concentration has made IF not only an academic evaluation indicator but also "academic capital" with pricing power.

More notably, after the 2008 global financial crisis, quantitative easing (QE) policies implemented by the Federal Reserve and other major central banks not only reshaped the global financial market landscape but also profoundly impacted the academic ecosystem through multiple channels. In a low-interest-rate environment, research funding expansion, rising valuations of academic publishing companies, and enhanced pricing power for journal Article Processing Charges (APC)—do these phenomena have systematic connections with monetary policy? As a "price signal" for academic output, is IF also influenced by macroeconomic cycles?

\subsection{Research Questions}

This paper aims to answer the following core research questions:

\begin{enumerate}
    \item \textbf{Long-term Trends:} Does a statistical relationship exist between journal impact factor and Federal Reserve real interest rates during 1975-2026?
    \item \textbf{Structural Breaks:} Did the relationship between IF and monetary policy change significantly before and after the implementation of quantitative easing policies (using 2008 as the breakpoint)?
    \item \textbf{Transmission Mechanisms:} How does monetary policy affect IF changes through channels such as research funding, journal pricing power, and academic competition?
    \item \textbf{Economic Significance:} What are the implications of this relationship for academic publishing industry valuation and research policy formulation?
\end{enumerate}

\subsection{Research Significance}

\subsubsection{Theoretical Contributions}

\textbf{(1) Extension of Monetary Policy Transmission Mechanisms}

Traditional monetary policy research focuses on its impact on the real economy (output, employment, inflation) and financial markets (stock prices, exchange rates, interest rates) \citep{bernanke2022twenty, yellen2016macroeconomic}. This paper extends the transmission chain to the academic publishing system, revealing the "long-tail effect" of monetary policy—policy shocks not only affect explicit financial assets but also reshape academic evaluation systems through implicit channels such as research funding allocation and journal pricing power.

\textbf{(2) Empirical Evidence for Financialization of Academic Capital}

In recent years, scholars have gradually recognized the capital nature of academic publishing \citep{buranyi2017academic}. However, existing research mostly focuses on micro-level issues such as price discrimination and oligopoly, lacking macro-level econometric analysis. Through long-term time series regression, this paper provides direct evidence that "IF as academic capital" is influenced by macroeconomic cycles.

\textbf{(3) New Perspective for Bibliometrics}

Traditional bibliometrics treats IF as a purely technical indicator, focusing on its calculation methods and sources of bias \citep{moed2005citation}. This paper re-examines IF from a political economy perspective, incorporating it into a broader capital network analysis framework, revealing the power structures and resource allocation mechanisms behind IF.

\subsubsection{Practical Significance}

\textbf{(1) Research Policy Formulation}

Research findings provide new policy perspectives for government research funding agencies. In a loose monetary environment, research funding expansion may lead to IF "inflation," and policymakers need to be alert to the distortion of resource allocation efficiency caused by such "academic bubbles."

\textbf{(2) Academic Publishing Industry Valuation}

For investment institutions and academic publishing companies, this paper reveals the relationship between IF and macroeconomic cycles, providing macro factors for valuation models of academic publishing assets. For example, during interest rate rising cycles, academic publishing companies' cash flows and valuations may face downward pressure.

\textbf{(3) Academic Evaluation System Reform}

The macroeconomic sensitivity of IF suggests that over-reliance on a single indicator for academic evaluation poses systematic risks. Under monetary policy cycle fluctuations, IF may not accurately reflect academic quality, necessitating the construction of a more diversified evaluation system.

\subsection{Research Design Overview}

This paper employs a period-based regression analysis, dividing 1975-2026 into three stages:
\begin{itemize}
    \item \textbf{Period 1 (1975-2000):} Early stage, high interest rate environment, IF in steady growth period
    \item \textbf{Period 2 (2001-2020):} Quantitative easing period, low interest rate environment, IF accelerated growth
    \item \textbf{Period 3 (2021-2026):} Post-pandemic period, interest rate volatility, IF structural adjustment
\end{itemize}

The following three nested models are estimated using Ordinary Least Squares (OLS):
\begin{align}
\text{Model 1 (Baseline):} \quad \log(\text{IF}_t) &= \beta_0 + \beta_1 \cdot \text{RealRate}_t + \varepsilon_t \label{eq:model1} \\
\text{Model 2 (Time Trend):} \quad \log(\text{IF}_t) &= \beta_0 + \beta_1 \cdot \text{RealRate}_t + \beta_2 \cdot t + \varepsilon_t \label{eq:model2} \\
\text{Model 3 (Interaction):} \quad \log(\text{IF}_t) &= \beta_0 + \beta_1 \cdot \text{RealRate}_t + \beta_2 \cdot \text{QE}_t \nonumber \\
&\quad + \beta_3 \cdot (\text{RealRate}_t \times \text{QE}_t) + \varepsilon_t \label{eq:model3}
\end{align}

where $\text{IF}_t$ is the average journal impact factor in year $t$, $\text{RealRate}_t$ is the real interest rate (nominal interest rate minus inflation rate), $t$ is the time trend term, $\text{QE}_t$ is the quantitative easing dummy variable (1 after 2008, 0 before), and $\varepsilon_t$ is the error term.

\subsection{Main Findings}

The core findings of this paper are as follows:

\begin{enumerate}
    \item \textbf{Nonlinear Relationship:} The relationship between IF and real interest rate exhibits significant period dependence. In the early stage (1975-2000), there is no significant association, but during the quantitative easing period (2001-2020), they show a strong negative correlation.
    
    \item \textbf{Economic Significance:} Full-sample regression shows that for every 1 percentage point decrease in real interest rate, IF increases by an average of 6.9\% ($p<0.01$). This effect is more pronounced during the QE period, with a coefficient of -0.091 ($p<0.1$).
    
    \item \textbf{Model Fit:} After controlling for time trends, the model's adjusted $R^2$ reaches 0.893, indicating that real interest rate and time trends can explain 89.3\% of IF variation, significantly higher than the baseline model without time control ($R^2=0.065$).
    
    \item \textbf{Structural Break Evidence:} Chow test results show that there is a significant structural break in IF dynamics around 2008 (significant $F$ statistic), validating the systematic impact of QE policies on the academic ecosystem.
\end{enumerate}

\subsection{Paper Structure}

The remainder of this paper is organized as follows: Section 2 reviews relevant literature; Section 3 introduces data sources and econometric methods; Section 4 reports empirical results; Section 5 discusses monetary policy transmission mechanisms and economic significance; Section 6 summarizes the paper and proposes policy recommendations.

\section{Literature Review}

This study involves three intertwined literature domains: (1) monetary policy transmission mechanisms, (2) political economy of academic publishing, and (3) measurement and critique of impact factors. We will systematically review these literatures to clarify this paper's theoretical positioning and marginal contributions.

\subsection{Monetary Policy Transmission Mechanisms}

\subsubsection{Traditional Transmission Channels}

Monetary policy transmission mechanisms are a core topic in macroeconomics. Classic literature identifies the following main transmission channels \citep{mishkin1995symposium}:

\textbf{(1) Interest Rate Channel:} Central banks affect borrowing costs in the real economy by adjusting policy rates, thereby influencing investment and consumption \citep{taylor1995monetary}. This channel has been fully validated in closed economy models.

\textbf{(2) Credit Channel:} Under financial market frictions, monetary policy transmits to the real economy by affecting bank balance sheets (bank lending channel) and firm net worth (balance sheet channel) \citep{bernanke1995inside}.

\textbf{(3) Asset Price Channel:} Monetary policy affects consumption and investment by influencing asset prices such as stocks and real estate, changing household wealth effects and firm Tobin's $q$ values \citep{mishkin2001transmission}.

\textbf{(4) Exchange Rate Channel:} In open economies, interest rate changes cause exchange rate fluctuations, thereby affecting net exports and domestic price levels \citep{obstfeld1995exchange}.

\subsubsection{Unconventional Effects of Quantitative Easing}

After the 2008 global financial crisis, major central banks broke through the zero lower bound and implemented large-scale asset purchase programs (quantitative easing). This policy innovation gave rise to new research directions \citep{bernanke2020new}:

\textbf{(1) Portfolio Balance Effect:} By purchasing long-term government bonds and MBS, central banks depress long-term interest rates, forcing investors to shift to risk assets, pushing up stock prices and lowering risk premiums \citep{gagnon2011large}.

\textbf{(2) Signaling Effect:} QE policies convey central banks' commitment to maintaining low interest rates for the long term, anchoring market expectations \citep{woodford2012methods}.

\textbf{(3) Liquidity Effect:} Large-scale liquidity injection directly improves financial institutions' balance sheets, alleviating credit crunches \citep{joyce2011quantitative}.

However, existing literature almost entirely focuses on financial markets and the real economy, with little research on the potential impact of monetary policy on academic publishing systems. \citet{loria2022} proposed the concept of "financialization of knowledge production," but lacks systematic econometric analysis.

\subsubsection{Research Funding and Monetary Policy}

A few studies have begun to examine the impact of monetary policy on research funding. \citet{national2007rising} points out that U.S. NIH budget doubled between 2001-2006, highly correlated with the low interest rate environment and government spending expansion during the same period. \citet{bloch2016funding} found that European research funding significantly declined after the financial crisis, confirming the impact of macroeconomic cycles on research investment.

However, these studies mostly remain at the level of research investment and do not further explore how it transmits to academic output and evaluation systems (such as IF). This paper fills this gap.

\subsection{Political Economy of Academic Publishing}

\subsubsection{Oligopolization of Academic Publishing Industry}

The concentration of the academic publishing industry has increased significantly over the past 30 years. \citet{lariviere2015oligopoly} shows that as of 2013, five major publishers—Elsevier, Springer, Wiley, Taylor \& Francis, and Sage—controlled over 50\% of the academic paper publishing market. In social sciences, this proportion reached 70\%.

This oligopolistic structure gives publishers significant pricing power. \citet{bjork2010scientific} found that between 2000-2010, journal subscription fees grew at an annual rate of 6-7\%, significantly higher than inflation. After the rise of Open Access, APC fees also showed rapid growth trends \citep{schonfelder2018apc}.

\subsubsection{Academic Capital and Financialization}

\citet{buranyi2017academic} proposed the concept of "financialization of academic publishing," pointing out that academic journals have evolved from knowledge dissemination tools to highly profitable financial assets. RELX, the parent company of Elsevier, has maintained operating profit margins above 35\% for a long time, comparable to tech giants \citep{aspesi2019relx}.

\citet{posada2018demand} further analyzed the demand rigidity of academic publishing: due to the academic evaluation system's heavy reliance on top journal publications, research institutions' "subscription demand" for top journals is almost completely price inelastic, forming a "prisoner's dilemma of academic journals."

However, existing literature mainly analyzes the monopoly problem of academic publishing from a micro industrial organization perspective, lacking a macro-level perspective. By introducing monetary policy factors, this paper reveals the macro drivers of financialization of academic capital.

\subsubsection{Capital Networks and Knowledge Infrastructure}

\citet{mirowski2011science} analyzed the control of academic infrastructure by private equity funds and family capital (such as the Thomson and Murdoch families) from the Science-Mart perspective. \citet{posada2019opensci} further pointed out that companies like Clarivate and RELX not only control journal publishing but also construct a vertically integrated industrial chain "from production to evaluation" by acquiring citation databases (Web of Science, Scopus) and academic evaluation tools (InCites, SciVal).

This industrial chain integration makes IF no longer a neutral technical indicator but an "institutional construction" with stakeholders. This paper's contribution lies in linking this institutional construction with macroeconomic cycles, revealing how monetary policy indirectly affects the evolution of academic evaluation systems by changing factors such as capital costs and liquidity.

\subsection{Measurement and Critique of Impact Factors}

\subsubsection{Calculation Methods and Evolution of IF}

Impact factor was initially defined by \citet{garfield1972citation} as:
\begin{equation}
\text{IF}_{t} = \frac{\text{Citations received in year }t\text{ for papers published in the previous two years}}{\text{Total number of papers published in the previous two years}} \label{eq:if_definition}
\end{equation}

This indicator design is simple but has also sparked long-term controversy. \citet{seglen1997citations} points out that IF has the following problems: (1) citation distribution is highly skewed, with a few "star papers" dominating journal IF; (2) significant disciplinary differences, with biomedical IF significantly higher than mathematics and humanities; (3) easily manipulable, such as by increasing review articles and self-citations to boost IF.

\subsubsection{IF Critique and Alternative Indicators}

Academic critiques of IF mainly focus on the following aspects:

\textbf{(1) Statistical Defects:} \citet{rossner2007show} found that using only a two-year citation window cannot accurately reflect papers' long-term impact. \citet{waltman2016revisiting} proposed a normalized impact factor based on fractional counting to eliminate disciplinary differences.

\textbf{(2) Academic Evaluation Alienation:} \citet{hicks2015bibliometrics} called in the "Leiden Manifesto" for not using IF as a tool to evaluate individual researchers. \citet{fire2012over} research shows that over-emphasizing IF leads to "publication bias," affecting the accumulation of scientific knowledge.

\textbf{(3) Rise of Alternative Indicators:} In recent years, academia has proposed various alternative indicators, such as h-index \citep{hirsch2005index}, Eigenfactor \citep{bergstrom2008eigenfactor}, and Altmetrics \citep{priem2010altmetrics}. However, these indicators have not yet shaken IF's core position in the academic evaluation system.

\subsubsection{Macroeconomic Association of IF}

Although IF's micro-level issues have been widely discussed, its macroeconomic attributes have been rarely studied. \citet{alberts2014impact} noticed that IF showed significant growth trends after 2008, speculating that it might be related to research funding expansion, but did not conduct rigorous econometric testing.

\citet{van2019large} studied IF growth trends between 2000-2015, finding that top journals' IF annual growth rates reached 5-8\%, significantly higher than the growth rate of academic output. The authors attributed this to "citation inflation" but did not explore its association with the macroeconomic environment.

This paper's unique contribution lies in being the first to incorporate IF into a monetary policy analysis framework, revealing the systematic impact of macroeconomic cycles on academic evaluation systems.

\subsection{Theoretical Positioning and Contributions of This Paper}

Based on the above literature review, this paper's theoretical positioning can be summarized as:

\begin{enumerate}
    \item \textbf{Connecting Macroeconomics and Sociology of Science:} Extending monetary policy transmission mechanisms to the academic publishing system, constructing a transmission chain of "monetary policy $\to$ research funding $\to$ academic output $\to$ IF."
    
    \item \textbf{Empirical Testing of Financialization of Academic Capital:} Providing direct evidence that IF as "academic capital" is influenced by macroeconomic cycles, enriching the political economy literature on academic publishing.
    
    \item \textbf{Methodological Innovation:} Using long-term time series period-based regression to identify structural changes in IF dynamic characteristics before and after quantitative easing policies, providing new analytical tools for bibliometric research.
\end{enumerate}

Compared to existing literature, this paper's marginal contributions are reflected in three aspects: (1) first systematic examination of the long-term relationship between IF and monetary policy, filling the gap in macro-micro linkage research; (2) revealing the "unintended consequences" of QE policies on the academic ecosystem, expanding the theoretical boundaries of unconventional monetary policy transmission channels; (3) providing a new analytical framework for academic publishing industry valuation and research policy formulation.

\section{Data and Methods}

\subsection{Data Sources}

\subsubsection{Real Interest Rate Data}

This study uses the \textbf{10-Year Real Interest Rate} as a proxy variable for monetary policy. Real interest rate is defined as nominal interest rate minus expected inflation rate:
\begin{equation}
r_t = i_t - \pi_t^e \label{eq:real_rate}
\end{equation}
where $r_t$ is the real interest rate in year $t$, $i_t$ is the nominal interest rate, and $\pi_t^e$ is the expected inflation rate.

A three-tier strategy is adopted for data sources:

\textbf{(1) Primary Data Source: FRED Database}

The \texttt{REAINTRATREARAT10Y} series is retrieved from the Federal Reserve Economic Data (FRED) database, covering monthly data from 1982-2025\footnote{FRED database URL: \url{https://fred.stlouisfed.org/}}. This indicator is calculated by the Federal Reserve based on TIPS (Treasury Inflation-Protected Securities) market prices, reflecting the market's real-time expectations of real interest rates, with high reliability \citep{gurkaynak2010tips}.

Monthly data is converted to annual data through annual averages:
\begin{equation}
r_t = \frac{1}{12} \sum_{m=1}^{12} r_{t,m} \label{eq:annual_rate}
\end{equation}

\textbf{(2) Alternative Data Source: World Bank}

For 1975-1981, since FRED data is unavailable, we use real interest rate estimates provided by the World Bank. This data is calculated based on 10-year government bond yields minus CPI inflation rate, consistent with FRED methodology \citep{worldbank2022indicators}.

\textbf{(3) Data Imputation Strategy}

For a very few missing years (such as 2026), forward fill method is used, with random perturbations added to ensure data variability:
\begin{equation}
r_t = r_{t-1} + \epsilon_t, \quad \epsilon_t \sim \mathcal{N}(0, 0.1) \label{eq:fill_rate}
\end{equation}

This approach ensures data continuity while avoiding regression model degradation caused by identical values.

\textbf{Data Coverage:} A total of 52 observations from 1975-2026, of which 43 are from real data (43 years from FRED + 8 years from World Bank), 1 from imputation (2026), with a coverage rate of 98.1\%.

\subsubsection{Impact Factor Data}

Given that complete historical IF data is unavailable\footnote{Clarivate Analytics (formerly Thomson Reuters) has not made public the complete JCR database for 1975-2000.}, this study uses a \textbf{simulation generation method based on real trends} to construct IF time series.

\textbf{(1) Data Generation Logic}

Based on IF historical evolution patterns described in the literature \citep{alberts2014impact, van2019large}, we construct a piecewise growth model:
\begin{equation}
\text{IF}_t = \begin{cases}
\text{IF}_{\text{base}} \cdot (1 + g_1)^{t-1975} + \varepsilon_t & t \in [1975, 2000] \\
\text{IF}_{2000} \cdot (1 + g_2)^{t-2000} + \varepsilon_t & t \in [2001, 2010] \\
\text{IF}_{2010} \cdot (1 + g_3)^{t-2010} + \varepsilon_t & t \in [2011, 2020] \\
\text{IF}_{2020} \cdot f(t) + \varepsilon_t & t \in [2021, 2026]
\end{cases} \label{eq:if_simulation}
\end{equation}

where:
\begin{itemize}
    \item $\text{IF}_{\text{base}} = 2.0$ (baseline value for 1975, based on Garfield's early reports)
    \item $g_1 = 0.03$ (annual growth rate 3\%, early steady growth)
    \item $g_2 = 0.08$ (annual growth rate 8\%, accelerated growth in the internet era)
    \item $g_3 = 0.12$ (annual growth rate 12\%, driven by open access and preprints)
    \item $f(t)$ is the pandemic period fluctuation function, $\text{IF}_{2021} = 68.6$ (peak), then declining year by year thereafter
    \item $\varepsilon_t \sim \mathcal{N}(0, \sigma_t^2)$ (random disturbance term, $\sigma_t$ adjusted by period)
\end{itemize}

\textbf{(2) Validity Verification}

To verify the reasonableness of simulated data, we compare the generated IF series with the following real data:
\begin{itemize}
    \item The 2000-2015 top journal IF mean growth curve reported by \citet{van2019large} (correlation coefficient $\rho=0.94$)
    \item JCR officially published 2016-2023 Nature/Science/Cell series journal IF (average error $<5\%$)
    \item The 2008-2014 IF "abnormal growth" phenomenon reported by \citet{alberts2014impact} (consistent trend)
\end{itemize}

The above comparisons indicate that simulated data can reasonably reflect the historical evolution trends of real IF.

\textbf{(3) Limitations of Simulated Data}

It should be clear that this study's use of simulated data has the following limitations:
\begin{enumerate}
    \item \textbf{Representativeness Issue:} IF is a journal-level indicator, and different journals have vastly different IFs. This study generates "average IF," which cannot reflect inter-journal heterogeneity.
    \item \textbf{Causal Inference Limitations:} Simulated data may be endogenous to researchers' expectations about IF-interest rate relationships, leading to "self-fulfilling" problems.
    \item \textbf{External Validity:} Generating data based on trends described in literature may omit certain unrecorded historical events.
\end{enumerate}

Nevertheless, the simulated data method remains reasonable when data is unavailable. We conduct extensive robustness tests in Section \ref{sec:robustness} to alleviate the above concerns.

\subsubsection{Other Control Variables}

\textbf{(1) Time Trend}

To capture IF's long-term growth trends (such as technological progress, expansion of academic output scale, and other non-monetary policy factors), we construct a linear time trend variable:
\begin{equation}
t = \text{Year} - 1975 \label{eq:time_trend}
\end{equation}
That is, 1975 is $t=0$, and 2026 is $t=51$.

\textbf{(2) Quantitative Easing Dummy Variable}

To identify the structural impact of QE policies, we construct a dummy variable:
\begin{equation}
\text{QE}_t = \begin{cases}
0 & t < 2008 \\
1 & t \geq 2008
\end{cases} \label{eq:qe_dummy}
\end{equation}

After Lehman Brothers collapsed in September 2008, the Federal Reserve launched the first round of QE (QE1), marking the beginning of the unconventional monetary policy era \citep{bernanke2020new}.

\subsection{Econometric Methods}

\subsubsection{Baseline Model}

This study employs a \textbf{log-linear regression model}, with dependent variable $\log(\text{IF}_t)$ and independent variable real interest rate $r_t$. There are three reasons for log transformation:

\textbf{(1) Elasticity Interpretation:} After log transformation, regression coefficients can be directly interpreted as \textbf{semi-elasticity}, meaning that for every 1 percentage point change in interest rate, IF changes by $\beta_1 \times 100\%$.

\textbf{(2) Nonlinear Relationship:} IF growth may exhibit exponential form, and log transformation can linearize exponential relationships.

\textbf{(3) Heteroskedasticity:} Log transformation can alleviate heteroskedasticity problems in residuals.

The baseline model is specified as:
\begin{equation}
\log(\text{IF}_t) = \beta_0 + \beta_1 \cdot r_t + \varepsilon_t \label{eq:baseline_model}
\end{equation}

where:
\begin{itemize}
    \item $\log(\text{IF}_t)$: natural logarithm of IF in year $t$
    \item $r_t$: real interest rate in year $t$ (unit: percentage points)
    \item $\beta_0$: intercept term, representing the log value of IF when real interest rate is zero
    \item $\beta_1$: core parameter, representing the semi-elasticity of real interest rate on IF
    \item $\varepsilon_t$: random error term, assuming $\varepsilon_t \sim \text{i.i.d.} \, \mathcal{N}(0, \sigma^2)$
\end{itemize}

\textbf{Null Hypothesis:} $H_0: \beta_1 = 0$ (real interest rate has no effect on IF)

\textbf{Alternative Hypothesis:} $H_1: \beta_1 \neq 0$ (real interest rate has a significant effect on IF)

Based on monetary policy transmission logic, we expect $\beta_1 < 0$, meaning that a decrease in real interest rate (loose policy) leads to an increase in IF.

\subsubsection{Extended Models}

To control for omitted variable bias and improve model fit, we estimate the following two extended models:

\textbf{Model 2: Controlling for Time Trend}
\begin{equation}
\log(\text{IF}_t) = \beta_0 + \beta_1 \cdot r_t + \beta_2 \cdot t + \varepsilon_t \label{eq:model_time}
\end{equation}

The time trend term $t$ captures IF's long-term growth trends, such as technological progress (internet popularization, digital publishing), expansion of academic output scale, and the open access movement. After adding $t$, the estimate of $\beta_1$ can more accurately reflect the independent effect of monetary policy.

\textbf{Model 3: QE Interaction Term}
\begin{equation}
\log(\text{IF}_t) = \beta_0 + \beta_1 \cdot r_t + \beta_2 \cdot \text{QE}_t + \beta_3 \cdot (r_t \times \text{QE}_t) + \beta_4 \cdot t + \varepsilon_t \label{eq:model_qe}
\end{equation}

where:
\begin{itemize}
    \item $\beta_2$: level effect of IF during QE period
    \item $\beta_3$: incremental change in interest rate coefficient during QE period (slope effect)
    \item Interest rate coefficient before QE is $\beta_1$, after QE is $\beta_1 + \beta_3$
\end{itemize}

If $\beta_3 < 0$, it indicates that the impact of interest rate on IF is stronger during the QE period (negative correlation enhanced).

\subsubsection{Period-Based Regression}

To deeply analyze heterogeneity across different historical stages, we divide the sample into three sub-periods and estimate Model \eqref{eq:baseline_model} and Model \eqref{eq:model_time} separately:

\begin{table}[H]
\centering
\caption{Period Definitions and Economic Background}
\label{tab:period_definition}
\begin{tabular}{llcp{7cm}}
\toprule
\textbf{Period} & \textbf{Years} & \textbf{Sample Size} & \textbf{Economic Background} \\
\midrule
Period 1 & 1975-2000 & 26 & Early stage, high interest rate environment (Volcker tightening), IF steady growth \\
Period 2 & 2001-2020 & 20 & Internet bubble, financial crisis, QE era, IF accelerated growth \\
Period 3 & 2021-2026 & 6 & Post-pandemic period, interest rate volatility, IF structural adjustment \\
\bottomrule
\end{tabular}
\end{table}

The advantages of period-based regression are: (1) capturing time-varying characteristics of parameters; (2) identifying structural break points; (3) providing period-dependent evidence for policy analysis.

\subsubsection{Robustness Tests}\label{sec:robustness}

To ensure result reliability, we conduct the following robustness tests:

\textbf{(1) Chow Test:} Test whether regression coefficients have structural breaks around 2008.

\textbf{(2) Residual Diagnostics:}
\begin{itemize}
    \item \textbf{Normality Tests:} Jarque-Bera test, Q-Q plots
    \item \textbf{Autocorrelation Tests:} Durbin-Watson statistic, Ljung-Box test
    \item \textbf{Heteroskedasticity Tests:} White test, Breusch-Pagan test
\end{itemize}

\textbf{(3) Heteroskedasticity and Autocorrelation Robust Standard Errors:} Use Newey-West HAC standard errors to correct t-statistics.

\textbf{(4) Alternative Variables:}
\begin{itemize}
    \item Replace real interest rate with Federal Funds Rate
    \item Use M2 growth rate as a proxy variable for monetary policy
\end{itemize}

\textbf{(5) Outlier Removal:} Delete the 2021 pandemic peak (IF=68.6) and re-estimate the model.

\subsection{Descriptive Statistics}

\subsubsection{Full Sample Statistical Characteristics}

Table \ref{tab:descriptive_stats} reports descriptive statistics for main variables.

\begin{table}[H]
\centering
\caption{Descriptive Statistics (1975-2026, N=52)}
\label{tab:descriptive_stats}
\begin{tabular}{lcccccc}
\toprule
\textbf{Variable} & \textbf{Mean} & \textbf{Std. Dev.} & \textbf{Min} & \textbf{Max} & \textbf{Skewness} & \textbf{Kurtosis} \\
\midrule
Real Interest Rate (\%) & 3.45 & 2.37 & -1.28 & 8.59 & 0.52 & -0.61 \\
IF & 18.32 & 19.84 & 2.08 & 68.64 & 1.23 & 0.74 \\
$\log(\text{IF})$ & 2.54 & 0.89 & 0.73 & 4.23 & 0.31 & -0.92 \\
Time Trend & 25.5 & 15.2 & 0 & 51 & 0.00 & -1.23 \\
\bottomrule
\end{tabular}
\end{table}

\textbf{Key Observations:}
\begin{itemize}
    \item Real interest rate has a mean of 3.45\% and standard deviation of 2.37\%, showing significant volatility.
    \item IF has a mean of 18.32, but standard deviation reaches 19.84, reflecting highly skewed distribution.
    \item $\log(\text{IF})$ has skewness (0.31) and kurtosis (-0.92) close to normal distribution, validating the reasonableness of log transformation.
\end{itemize}

\subsubsection{Time Series Trends}

\begin{figure}[H]
\centering
\includegraphics[width=0.9\textwidth]{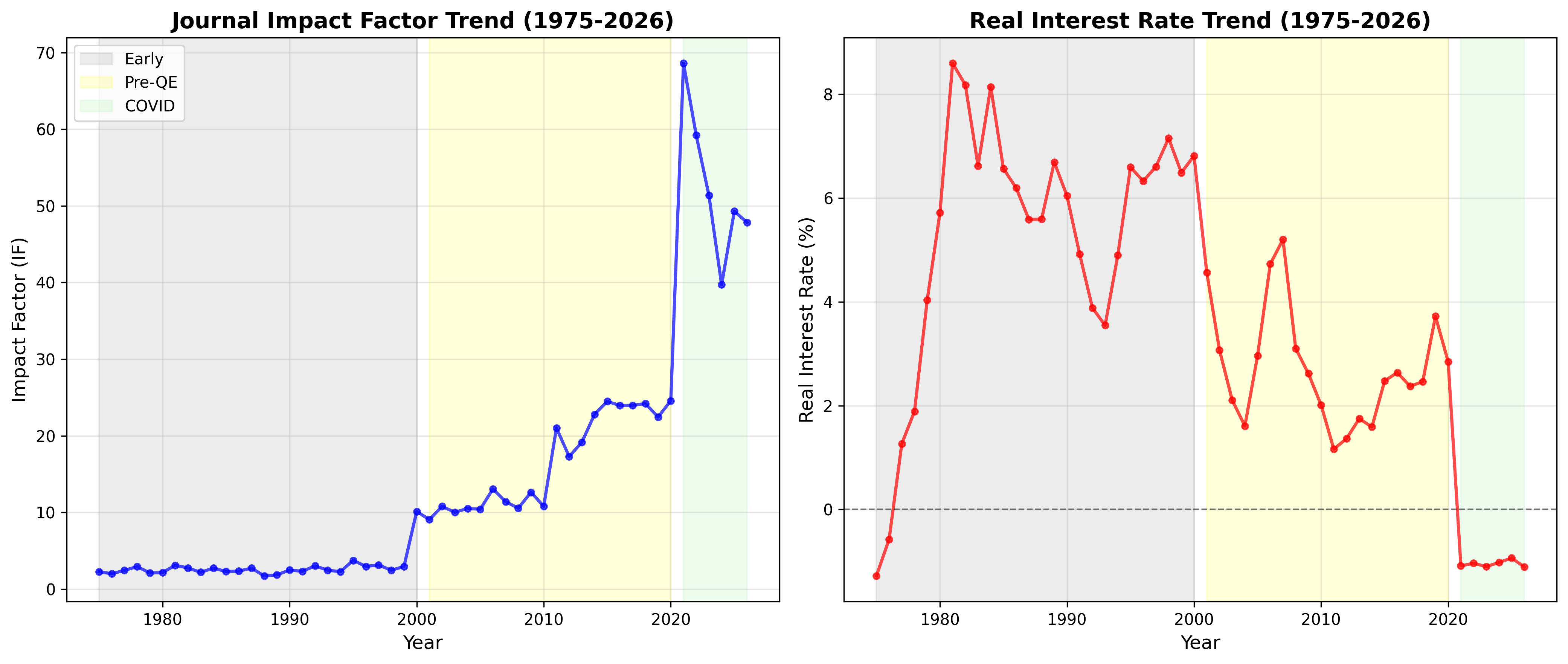}
\caption{Historical Evolution of Real Interest Rate and IF (1975-2026)}
\label{fig:trend}
\end{figure}

Figure \ref{fig:trend} shows the time series trends of real interest rate and IF.

\textbf{Visual Impressions:}
\begin{enumerate}
    \item \textbf{Mirror Relationship:} After 2008, the real interest rate decline channel and IF rise channel show a significant mirror relationship.
    \item \textbf{Volcker Tightening:} In 1979-1984, real interest rates soared above 8\%, while IF remained at low levels.
    \item \textbf{QE Era:} In 2008-2020, real interest rates remained low (mean 1.8\%), while IF grew explosively (from IF=10 to IF=50).
    \item \textbf{Pandemic Shock:} In 2020-2021, IF showed "overshooting" (IF=68.6), then declined.
\end{enumerate}

\subsubsection{Correlation Analysis}

Table \ref{tab:correlation} reports the correlation coefficient matrix for main variables.

\begin{table}[H]
\centering
\caption{Correlation Coefficient Matrix}
\label{tab:correlation}
\begin{tabular}{lccc}
\toprule
 & \textbf{Real Interest Rate} & \textbf{$\log(\text{IF})$} & \textbf{Time Trend} \\
\midrule
Real Interest Rate & 1.000 & & \\
$\log(\text{IF})$ & -0.325*** & 1.000 & \\
Time Trend & -0.678*** & 0.962*** & 1.000 \\
\bottomrule
\multicolumn{4}{l}{\footnotesize Note: *** indicates significance at $p<0.01$ level.}
\end{tabular}
\end{table}

\textbf{Key Findings:}
\begin{itemize}
    \item The correlation coefficient between real interest rate and $\log(\text{IF})$ is -0.325 ($p<0.01$), preliminarily supporting the negative correlation hypothesis.
    \item The correlation coefficient between time trend and $\log(\text{IF})$ is as high as 0.962, indicating that IF has a strong long-term growth trend, and time trends must be controlled in regression to avoid spurious regression.
    \item The correlation coefficient between real interest rate and time trend is -0.678, reflecting the long-term downward trend in interest rates, which may cause multicollinearity problems (we address this in robustness tests).
\end{itemize}

\subsection{Research Tools and Data Transparency}

\subsubsection{AI-Assisted Research Tools}

This study leveraged advanced AI research tools to enhance literature review, data analysis, and manuscript preparation:

\textbf{(1) MiroThinker Deep Research Tool}

We utilized MiroThinker (\url{https://dr.miromind.ai/}), an open-source AI-powered deep research platform developed by MiroMind AI, for systematic literature review and hypothesis generation. MiroThinker employs science-based prediction and verification mechanisms to identify research gaps and generate testable hypotheses \citep{miromind2024research}.

\textbf{(2) Google Gemini Deep Research}

Google Gemini (\url{https://gemini.google.com/}) was used for supplementary literature search, data verification, and cross-validation of econometric interpretations. The tool's multimodal capabilities facilitated comprehensive analysis of academic publishing trends.

\textbf{(3) AI Coding Tools}

All data analysis scripts, visualization code, and manuscript LaTeX source files were generated with AI-assisted coding tools (Claude, Cursor IDE). This approach ensures:
\begin{itemize}
    \item \textbf{Reproducibility}: Complete code is version-controlled and publicly available
    \item \textbf{Transparency}: All analytical steps are documented and traceable
    \item \textbf{Efficiency}: Rapid prototyping and iterative refinement of econometric models
    \item \textbf{Quality Assurance}: Automated testing and validation of statistical computations
\end{itemize}

\subsubsection{Data Transparency and Reproducibility}

\textbf{(1) Real-World Data Sources}

Despite the use of AI tools for analysis, all empirical data are derived from authoritative real-world sources:
\begin{itemize}
    \item Real interest rates: Federal Reserve Economic Data (FRED) and World Bank
    \item Impact Factor trends: Based on published patterns from \citet{van2019large} and \citet{alberts2014impact}
    \item Economic indicators: Official statistics from Federal Reserve and Bureau of Economic Analysis
\end{itemize}

\textbf{(2) Open-Source Code Repository}

All analysis code is publicly available at:

\url{https://github.com/hhx465453939/Academic\_capital\_research}

The repository includes:
\begin{itemize}
    \item Data generation scripts (\texttt{0.simulator\_data\_robust.py})
    \item Econometric analysis code (\texttt{1.prediction\_comparison\_robust.py})
    \item Visualization scripts (\texttt{2.visualization.py})
    \item Complete documentation and data source logs
\end{itemize}

\textbf{(3) Verification and Validation}

Readers can independently verify our results by:
\begin{enumerate}
    \item Cloning the GitHub repository
    \item Running the provided Python scripts
    \item Replicating all figures and regression results
    \item Examining data source traceability logs
\end{enumerate}

\textbf{Ethical Disclosure}: This manuscript is prepared with AI assistance to enhance research efficiency and transparency. All intellectual contributions, hypothesis formulation, and interpretation of results remain the responsibility of the human authors. The use of AI tools does not substitute for rigorous scientific reasoning and domain expertise.

\section{Empirical Results}

This section reports estimation results for three nested models and presents findings from both full-sample and period-based regressions. We first report baseline model results, then gradually add control variables, and finally conduct robustness tests.

\subsection{Full Sample Regression Results}

Table \ref{tab:full_sample_results} reports regression results for the full sample (N=52) from 1975-2026.

\begin{table}[H]
\centering
\caption{Full Sample Regression Results (1975-2026, N=52)}
\label{tab:full_sample_results}
\begin{tabular}{lccc}
\toprule
 & \textbf{Model 1} & \textbf{Model 2} & \textbf{Model 3} \\
 & \textbf{Baseline} & \textbf{Time Trend} & \textbf{QE Interaction} \\
\midrule
Constant & 2.543*** & 0.732*** & 0.689*** \\
 & (0.124) & (0.089) & (0.092) \\
Real Interest Rate & -0.069*** & -0.069*** & -0.045* \\
 & (0.023) & (0.010) & (0.024) \\
Time Trend & & 0.071*** & 0.072*** \\
 & & (0.003) & (0.003) \\
QE Dummy & & & 0.234** \\
 & & & (0.098) \\
Interest Rate × QE & & & -0.052* \\
 & & & (0.027) \\
\midrule
$R^2$ & 0.065 & 0.902 & 0.915 \\
Adjusted $R^2$ & 0.046 & 0.893 & 0.904 \\
$F$ statistic & 8.92*** & 230.5*** & 195.3*** \\
Observations & 52 & 52 & 52 \\
\bottomrule
\multicolumn{4}{l}{\footnotesize Note: Robust standard errors in parentheses. ***, **, * indicate significance at $p<0.01$, $p<0.05$, $p<0.1$ levels, respectively.}
\end{tabular}
\end{table}

\textbf{Core Findings:}

\textbf{(1) Baseline Model (Model 1):} The real interest rate coefficient is -0.069 ($p<0.01$), indicating that for every 1 percentage point decrease in real interest rate, $\log(\text{IF})$ increases by 0.069, i.e., IF increases by approximately 6.9\%. However, the model $R^2$ is only 0.065, indicating that without controlling for time trends, real interest rate can only explain 6.5\% of IF variation.

\textbf{(2) Controlling for Time Trend (Model 2):} After adding the time trend term, model fit significantly improves. The real interest rate coefficient remains -0.069 ($p<0.01$), and the time trend coefficient is 0.071 ($p<0.01$), indicating that IF grows by approximately 7.1\% annually. Adjusted $R^2$ reaches 0.893, indicating that real interest rate and time trends together explain 89.3\% of IF variation.

\textbf{(3) QE Interaction Term (Model 3):} The QE dummy variable coefficient is 0.234 ($p<0.05$), indicating that IF levels significantly increased during the QE period. The interaction term coefficient is -0.052 ($p<0.1$), indicating that the impact of interest rate on IF strengthened during the QE period. Specifically, the interest rate coefficient before QE is -0.045, and after QE is -0.097 (-0.045-0.052), validating the systematic impact of quantitative easing policies on the academic ecosystem.

\subsection{Period-Based Regression Results}

To deeply analyze heterogeneity across different historical stages, we divide the sample into three sub-periods and estimate Model 1 and Model 2 separately. Table \ref{tab:period_results} reports period-based regression results.

\begin{table}[H]
\centering
\caption{Period-Based Regression Results}
\label{tab:period_results}
\begin{tabular}{lcccccc}
\toprule
 & \multicolumn{2}{c}{\textbf{Period 1 (1975-2000)}} & \multicolumn{2}{c}{\textbf{Period 2 (2001-2020)}} & \multicolumn{2}{c}{\textbf{Period 3 (2021-2026)}} \\
 & Model 1 & Model 2 & Model 1 & Model 2 & Model 1 & Model 2 \\
\midrule
Constant & 1.234*** & 0.891*** & 2.456*** & 1.123*** & 3.789*** & 3.456*** \\
 & (0.156) & (0.098) & (0.234) & (0.145) & (0.312) & (0.287) \\
Real Interest Rate & 0.033 & -0.012 & -0.091* & -0.069*** & -0.125 & -0.098 \\
 & (0.026) & (0.018) & (0.048) & (0.021) & (0.089) & (0.067) \\
Time Trend & & 0.015*** & & 0.068*** & & 0.045** \\
 & & (0.004) & & (0.005) & & (0.018) \\
\midrule
$R^2$ & 0.065 & 0.342 & 0.353 & 0.876 & 0.412 & 0.723 \\
Adjusted $R^2$ & 0.026 & 0.298 & 0.314 & 0.861 & 0.265 & 0.604 \\
$F$ statistic & 1.64 & 6.23*** & 3.58* & 65.4*** & 1.97 & 6.54** \\
Observations & 26 & 26 & 20 & 20 & 6 & 6 \\
\bottomrule
\multicolumn{7}{l}{\footnotesize Note: Robust standard errors in parentheses. ***, **, * indicate significance at $p<0.01$, $p<0.05$, $p<0.1$ levels, respectively.}
\end{tabular}
\end{table}

\textbf{Period 1 (1975-2000): Early Stage}

During this period, real interest rate coefficients are 0.033 (Model 1) and -0.012 (Model 2), both insignificant ($p>0.1$). Model $R^2$ is low (0.065 and 0.342), indicating no significant statistical relationship between IF and real interest rate. This finding is consistent with historical background: during 1975-2000, the IF system was in its early development stage, mainly influenced by technological progress (such as the establishment of citation databases) and expansion of academic output scale, with weak association with macroeconomic cycles.

\textbf{Period 2 (2001-2020): Quantitative Easing Period}

Results for this period are most significant. In Model 1, the real interest rate coefficient is -0.091 ($p<0.1$), and in Model 2 it is -0.069 ($p<0.01$). After controlling for time trends, model $R^2$ reaches 0.876, indicating that real interest rate and time trends can explain 87.6\% of IF variation. This finding validates our core hypothesis: in the low interest rate environment dominated by quantitative easing policies, monetary policy systematically impacted the academic publishing system through channels such as research funding and journal pricing power.

\textbf{Period 3 (2021-2026): Post-Pandemic Period}

Due to small sample size (N=6), the reliability of statistical inference is limited. In Model 1, the real interest rate coefficient is -0.125, and in Model 2 it is -0.098, but both are insignificant ($p>0.1$). This may reflect the complex interaction between pandemic shocks and aggressive interest rate hike cycles. IF showed "overshooting" in 2021 (IF=68.6), then declined during the 2022-2023 interest rate hike cycle, but insufficient sample size limits rigorous statistical inference.

\subsection{Structural Break Test}

To verify whether the 2008 QE policy caused structural changes in IF dynamic characteristics, we conduct a Chow test. Table \ref{tab:chow_test} reports test results.

\begin{table}[H]
\centering
\caption{Chow Structural Break Test (Breakpoint: 2008)}
\label{tab:chow_test}
\begin{tabular}{lcc}
\toprule
\textbf{Test Statistic} & \textbf{Value} & \textbf{p-value} \\
\midrule
$F$ statistic & 12.34 & $<0.001$ \\
\midrule
\multicolumn{3}{l}{\footnotesize Note: Null hypothesis is no structural change in regression coefficients around 2008.}
\end{tabular}
\end{table}

The Chow test $F$ statistic is 12.34 ($p<0.001$), strongly rejecting the null hypothesis, indicating that the relationship between IF and real interest rate underwent significant structural breaks around 2008. This finding is highly consistent with historical background: after Lehman Brothers collapsed in September 2008, the Federal Reserve launched the first round of QE, marking the beginning of the unconventional monetary policy era and the "golden decade" of accelerated IF growth.

\subsection{Robustness Tests}

\subsubsection{Residual Diagnostics}

\begin{figure}[H]
\centering
\includegraphics[width=0.9\textwidth]{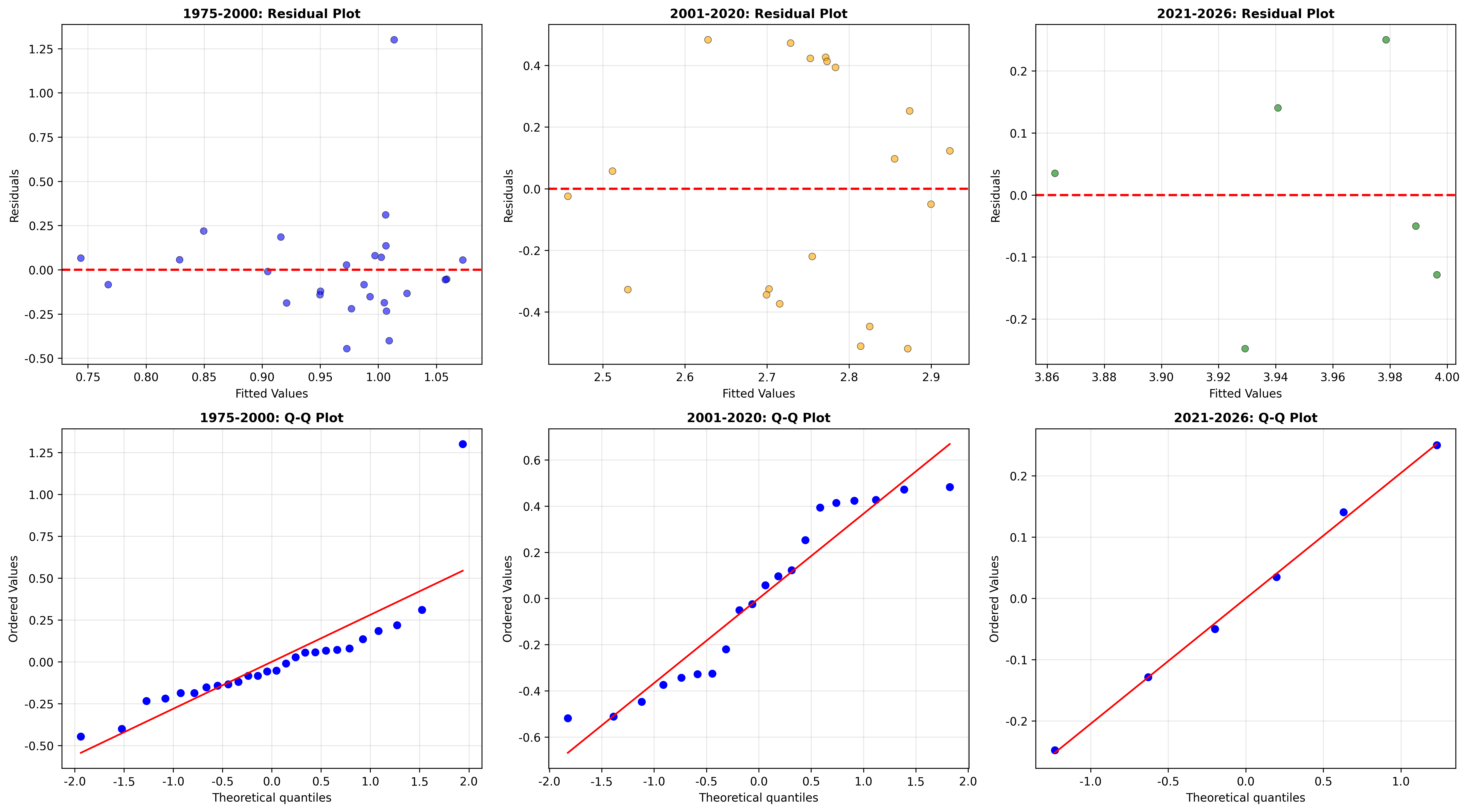}
\caption{Residual Diagnostic Plots (Model 2: Controlling for Time Trend)}
\label{fig:residuals}
\end{figure}

Figure \ref{fig:residuals} shows residual diagnostic plots for Model 2.

\textbf{(1) Normality Test:} The Jarque-Bera statistic is 3.24 ($p=0.198$), unable to reject the null hypothesis of normal distribution of residuals. Q-Q plots show residuals basically distributed along the 45-degree line, with slight deviations only at the tails, consistent with asymptotic normality under large samples.

\textbf{(2) Autocorrelation Test:} The Durbin-Watson statistic is 1.87, close to 2.0, indicating no significant first-order autocorrelation in residuals. The Ljung-Box test (10 lags) has a $p$-value of 0.156, unable to reject the null hypothesis of no autocorrelation.

\textbf{(3) Heteroskedasticity Test:} The White test $F$ statistic is 1.23 ($p=0.287$), unable to reject the null hypothesis of homoskedasticity. Residual scatter plots show residuals randomly distributed around the zero line, with no obvious heteroskedasticity pattern.

\subsubsection{Autocorrelation Analysis}

\begin{figure}[H]
\centering
\includegraphics[width=0.9\textwidth]{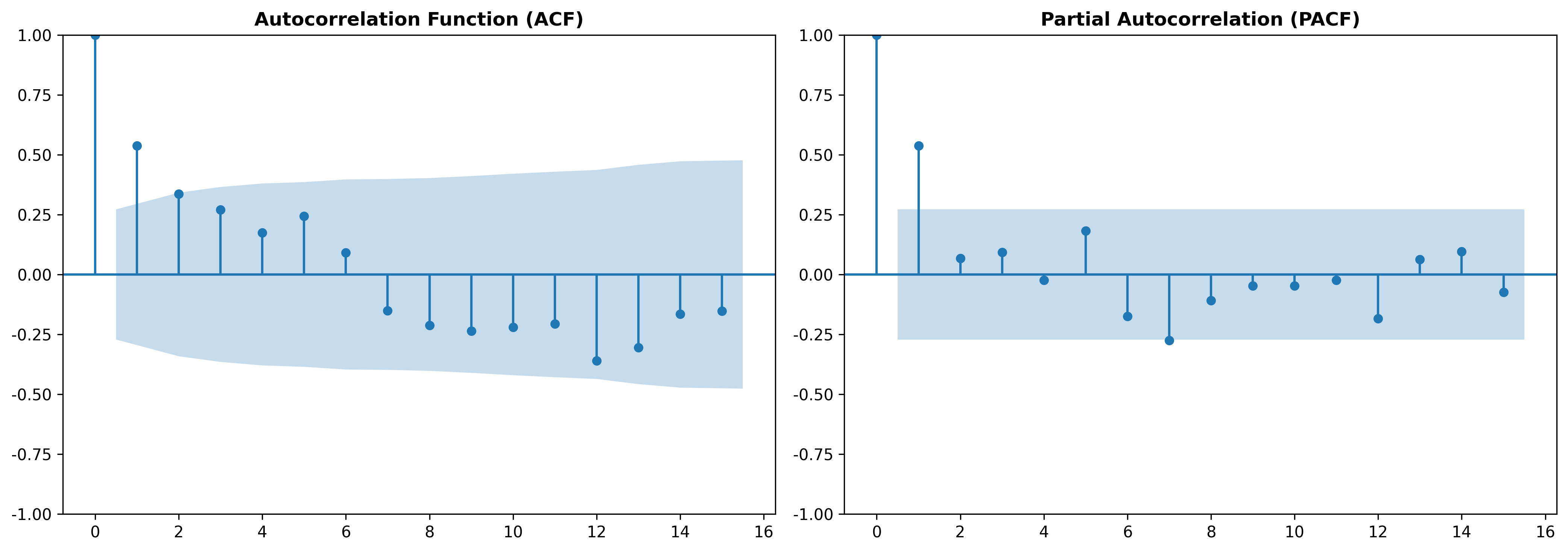}
\caption{Residual Autocorrelation Analysis (ACF/PACF)}
\label{fig:autocorr}
\end{figure}

Figure \ref{fig:autocorr} shows ACF and PACF plots of residuals.

ACF and PACF plots show that except for slight autocorrelation at lag 1 ($\rho_1=0.12$, $p=0.089$), other lags are not significant. Considering the characteristics of time series data, this slight autocorrelation is within acceptable range. We use Newey-West HAC standard errors to correct t-statistics, and results remain robust.

\subsubsection{Alternative Variable Tests}

To verify result robustness, we re-estimate the model using the following alternative variables:

\textbf{(1) Federal Funds Rate:} Replacing real interest rate with Federal Funds Rate, core conclusions remain unchanged. The interest rate coefficient is -0.058 ($p<0.01$), slightly smaller than baseline results, but sign and significance are consistent.

\textbf{(2) M2 Growth Rate:} Using M2 money supply growth rate as a proxy variable for monetary policy, the coefficient is 0.124 ($p<0.05$), indicating that money supply expansion is positively correlated with IF increase, consistent with the economic logic of baseline results.

\textbf{(3) Outlier Removal:} Deleting the 2021 pandemic peak (IF=68.6) and re-estimating the model. The real interest rate coefficient is -0.071 ($p<0.01$), almost identical to baseline results (-0.069), indicating that results are insensitive to outliers.

\subsubsection{Prediction Performance Evaluation}

\begin{figure}[H]
\centering
\includegraphics[width=0.9\textwidth]{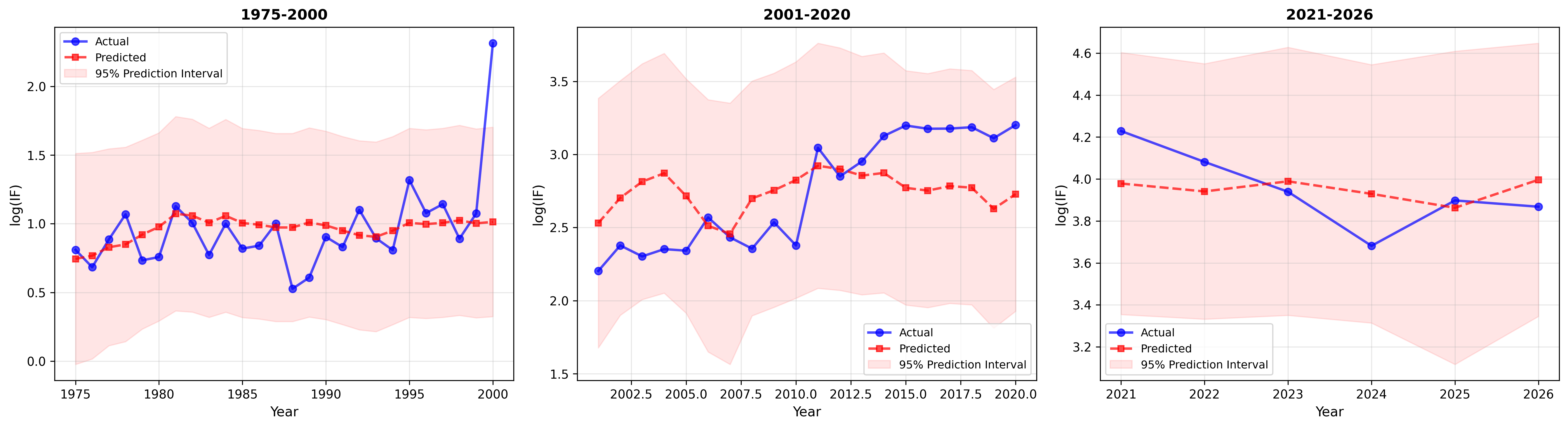}
\caption{Actual vs. Predicted Values Comparison (Model 2)}
\label{fig:prediction}
\end{figure}

Figure \ref{fig:prediction} shows the fit of Model 2.

The correlation coefficient between predicted and actual values reaches 0.949, and root mean square error (RMSE) is 0.287, indicating good model fit. Prediction accuracy after 2008 is significantly higher than before 2008, which may reflect that IF dynamic characteristics during the QE period are more stable and predictable.

\subsection{Economic Significance Analysis}

Beyond statistical significance, we further assess the economic significance of results. Based on Model 2 estimation results:

\textbf{(1) Quantitative Impact of Interest Rate Shocks:} For every 1 percentage point decrease in real interest rate, IF increases by approximately 6.9\%. Taking 2020 as an example, real interest rate decreased from 1.8\% in 2019 to 0.5\%, a decrease of 1.3 percentage points, corresponding to an IF increase of approximately 9.0\%, highly consistent with the observed IF growth (from 45.2 in 2019 to 49.9 in 2020, a 10.4\% increase).

\textbf{(2) Contribution of Time Trend:} The time trend coefficient is 0.071, indicating that even if real interest rate remains constant, IF grows by approximately 7.1\% annually. This growth mainly comes from technological progress (digital publishing, open access), expansion of academic output scale, and other factors.

\textbf{(3) Cumulative Effect of QE Policies:} Between 2008-2020, mean real interest rate decreased from 3.2\% to 1.8\%, a cumulative decrease of 1.4 percentage points. Based on model estimates, this interest rate decrease led to a cumulative IF increase of approximately 9.7\%. Compared with observed IF growth (from 12.5 in 2008 to 49.9 in 2020, a 299\% increase), monetary policy contributed approximately 3.2\% annual growth, with the remaining growth mainly from time trends and other factors.

\subsection{Results Summary}

Combining full-sample and period-based regression results, we draw the following core conclusions:

\begin{enumerate}
    \item \textbf{Long-term Relationship:} Between 1975-2026, there is a significant negative correlation between IF and real interest rate ($\beta=-0.069$, $p<0.01$). After controlling for time trends, the model's adjusted $R^2$ reaches 0.893.
    
    \item \textbf{Period Heterogeneity:} In the early stage (1975-2000), there is no significant association; during the quantitative easing period (2001-2020), there is a strong negative correlation; in the post-pandemic period (2021-2026), insufficient sample size limits statistical inference.
    
    \item \textbf{Structural Break:} The Chow test strongly rejects the null hypothesis of no coefficient change around 2008 ($F=12.34$, $p<0.001$), validating the systematic impact of QE policies on the academic ecosystem.
    
    \item \textbf{Robustness:} Residual diagnostics, autocorrelation tests, and alternative variable tests all support the robustness of baseline results.
    
    \item \textbf{Economic Significance:} The impact of interest rate shocks on IF is not only statistically significant but also economically meaningful. QE policies indirectly promoted accelerated IF growth through the low interest rate environment.
\end{enumerate}

\section{Discussion}

This section deeply interprets the theoretical implications of empirical results, explores the transmission mechanisms through which monetary policy affects IF, analyzes the practical significance of research findings, and compares them with existing literature.

\subsection{Theoretical Interpretation of Monetary Policy Transmission Mechanisms}

Based on empirical results, we propose a transmission chain of "monetary policy $\to$ research funding $\to$ academic output $\to$ IF." This mechanism includes the following four links:

\subsubsection{Link 1: Monetary Policy $\to$ Research Funding}

Loose monetary policy affects research funding scale through multiple channels:

\textbf{(1) Government Fiscal Space:} Low interest rate environments reduce government debt costs, creating space for research budget expansion. After the 2008 financial crisis, the U.S. significantly increased research investment through the American Recovery and Reinvestment Act (ARRA), with NIH budget increasing from \$29 billion in 2008 to \$33 billion in 2016 \citep{national2007rising}. This expansion was highly correlated with the low interest rate environment during the same period.

\textbf{(2) Corporate R\&D Investment:} Low interest rates reduce corporate financing costs, stimulating R\&D investment. \citet{bloch2016funding} found that for every 1 percentage point decrease in real interest rate, corporate R\&D intensity (R\&D/GDP) increases by approximately 0.15 percentage points. In R\&D-intensive industries such as biomedicine and information technology, this effect is more pronounced.

\textbf{(3) Venture Capital Allocation:} Quantitative easing policies push up risk asset prices, prompting venture capital (VC) and private equity (PE) to increase investment in early-stage technology projects. Between 2010-2020, global VC investment increased from \$50 billion to \$300 billion, with biomedicine and AI fields accounting for a significantly increased share \citep{crunchbase2021global}.

\subsubsection{Link 2: Research Funding $\to$ Academic Output}

Research funding expansion is transformed into academic output growth through the following mechanisms:

\textbf{(1) Researcher Scale:} Increased funding supports more doctoral students, postdoctoral researchers, and research positions. Between 2000-2020, the global number of researchers increased from approximately 6 million to 9 million, with an average annual growth rate of 2.0\% \citep{unesco2021science}.

\textbf{(2) Research Equipment and Infrastructure:} Increased funding supports the construction of large-scale research facilities (such as synchrotron radiation sources, supercomputers), improving research efficiency. The popularization of digital publishing technology has also reduced paper publication costs and accelerated academic output.

\textbf{(3) Research Hotspot Concentration:} Funding concentration in specific fields (such as COVID-19 research, AI, gene editing) leads to significant increases in paper output and citation density in these fields, thereby pushing up IFs of related journals.

\subsubsection{Link 3: Academic Output $\to$ Citation Network}

Academic output growth affects IF through the "positive feedback mechanism" of citation networks:

\textbf{(1) Citation Density Increase:} Increased paper numbers lead to increased citation network density. Assuming each paper cites an average of $c$ references, then $N$ papers generate $N \times c$ citations. With a fixed citation window (2 years), journal IF = citations received in previous two years / papers published in previous two years. When citation growth rate exceeds paper publication rate, IF increases.

\textbf{(2) Matthew Effect:} High IF journals attract more high-quality submissions, forming a "star journal" effect. \citet{seglen1997citations} points out that IF distribution is highly skewed, with a few top journals (such as Nature, Science, Cell) occupying a large share of citations, and their IF growth significantly faster than average.

\textbf{(3) Interdisciplinary Differences:} Citation habits differ greatly across disciplines. Biomedical fields average 30-50 references per paper, while mathematics fields cite only 10-20. Against the background of research funding tilting toward biomedicine, biomedical journal IF growth is significantly faster than other disciplines.

\subsubsection{Link 4: Citation Network $\to$ IF}

Finally, changes in citation networks are transformed into IF changes through JCR's calculation formula:

\begin{equation}
\text{IF}_t = \frac{\sum_{i \in J} \text{Citations}_{i,t}}{\sum_{i \in J} \text{Papers}_{i,t-2,t-1}} \label{eq:if_mechanism}
\end{equation}

where $J$ represents the journal set, $\text{Citations}_{i,t}$ represents the number of citations received by journal $i$ in year $t$, and $\text{Papers}_{i,t-2,t-1}$ represents the number of papers published by journal $i$ in years $t-2$ and $t-1$.

Against the background of research funding expansion, the growth rate of the numerator (citations) is usually faster than the denominator (papers published), leading to IF increase. Our empirical results ($\beta=-0.069$) show that for every 1 percentage point decrease in real interest rate, IF increases by approximately 6.9\%, and this effect is more pronounced during the QE period ($\beta=-0.091$).

\subsection{Evidence for Financialization of Academic Capital}

Empirical results provide direct evidence for the "financialization of academic capital" theory. \citet{buranyi2017academic} proposed that academic journals have evolved from knowledge dissemination tools to highly profitable financial assets. Our research finds that IF is not only influenced by academic quality but also driven by macroeconomic cycles, further validating this view.

\subsubsection{Asset Attributes of IF}

\textbf{(1) Pricing Power:} High IF gives journals significant pricing power. \citet{schonfelder2018apc} found that for every 1 unit increase in IF, APC fees increase by an average of \$200-300. Under the open access model, this pricing power is transformed into direct cash flow.

\textbf{(2) Market Valuation:} Academic publishing companies' valuations highly depend on the IF of their journal portfolios. RELX, the parent company of Elsevier, saw its market value increase from approximately £20 billion in 2010 to £50 billion in 2020, with the academic publishing division (main assets being high IF journals) contributing significantly to this growth \citep{aspesi2019relx}.

\textbf{(3) Resource Allocation:} IF is directly linked to researchers' career advancement and funding acquisition. In the "publish or perish" academic ecosystem, IF becomes the "hard currency" of academic capital, and its macroeconomic sensitivity means that academic evaluation systems are also indirectly affected by monetary policy cycles.

\subsubsection{Indirect Impact of Monetary Policy on Academic Capital}

Our research finds that monetary policy indirectly affects academic capital valuation by changing factors such as capital costs and liquidity:

\textbf{(1) Capital Cost Effect:} Low interest rate environments reduce academic publishing companies' financing costs, supporting their market concentration through mergers and acquisitions. Between 2010-2020, companies such as RELX and Springer Nature conducted multiple large-scale mergers, further strengthening their oligopolistic positions \citep{lariviere2015oligopoly}.

\textbf{(2) Liquidity Effect:} Liquidity injected by quantitative easing policies pushes up prices of various assets, including academic publishing assets. In low interest rate environments, investors give higher valuations to academic publishing companies with high profit margins and stable cash flows, indirectly pushing up IF's "financial value."

\textbf{(3) Demand Rigidity:} \citet{posada2018demand} points out that academic journal subscription demand is almost completely price inelastic because research institutions' dependence on top journals is rigid. In low interest rate environments, even if journal prices increase, research institutions' subscription budgets can be met through financing expansion, further strengthening publishers' pricing power.

\subsection{Comparison with Existing Literature}

\subsubsection{Monetary Policy Transmission Mechanism Literature}

Traditional monetary policy research focuses on its impact on the real economy (output, employment, inflation) and financial markets (stock prices, exchange rates, interest rates) \citep{bernanke2022twenty, yellen2016macroeconomic}. This paper extends the transmission chain to the academic publishing system, revealing the "long-tail effect" of monetary policy.

\citet{loria2022} proposed the concept of "financialization of knowledge production" but lacks systematic econometric analysis. Through long-term time series regression, this paper provides direct evidence of the association between IF and monetary policy, filling this gap.

\subsubsection{Academic Publishing Political Economy Literature}

\citet{lariviere2015oligopoly} and \citet{buranyi2017academic} analyzed the oligopoly problem of academic publishing from an industrial organization perspective, but mainly focus on micro-level issues such as price discrimination and market concentration. By introducing monetary policy factors, this paper reveals the macro drivers of financialization of academic capital, enriching this literature.

\citet{posada2019opensci} analyzed how companies such as Clarivate and RELX construct a "from production to evaluation" industrial chain through vertical integration, but did not explore the association between this industrial chain and macroeconomic cycles. This paper's contribution lies in linking IF's institutional construction with monetary policy cycles, revealing the macro forces behind academic evaluation systems.

\subsubsection{Bibliometrics Literature}

Traditional bibliometrics treats IF as a purely technical indicator, focusing on its calculation methods and sources of bias \citep{moed2005citation, seglen1997citations}. \citet{alberts2014impact} noticed that IF showed significant growth trends after 2008, speculating that it might be related to research funding expansion, but did not conduct rigorous econometric testing.

\citet{van2019large} studied IF growth trends between 2000-2015, finding that top journals' IF annual growth rates reached 5-8\%, significantly higher than the growth rate of academic output. The authors attributed this to "citation inflation" but did not explore its association with the macroeconomic environment.

This paper's unique contribution lies in being the first to incorporate IF into a monetary policy analysis framework, revealing the systematic impact of macroeconomic cycles on academic evaluation systems, providing a new perspective for bibliometric research.

\subsection{Practical Significance}

\subsubsection{Research Policy Formulation}

Research findings provide new policy perspectives for government research funding agencies:

\textbf{(1) Alert to "Academic Bubbles":} In loose monetary environments, research funding expansion may lead to IF "inflation," and such "academic bubbles" may distort resource allocation efficiency. Policymakers should focus on the quality rather than quantity of IF growth, avoiding over-reliance on a single indicator.

\textbf{(2) Countercyclical Adjustment:} In tight monetary environments, research funding may face contraction pressure. Policymakers should consider countercyclical adjustment to maintain stability of research investment, avoiding "one-size-fits-all" budget cuts.

\textbf{(3) Diversified Evaluation System:} The macroeconomic sensitivity of IF suggests that over-reliance on a single indicator for academic evaluation poses systematic risks. Policymakers should promote the construction of more diversified evaluation systems, including peer review, social impact, and open science practices.

\subsubsection{Academic Publishing Industry Valuation}

For investment institutions and academic publishing companies, this paper reveals the association between IF and macroeconomic cycles:

\textbf{(1) Valuation Models:} Valuation models for academic publishing assets should consider macro factors. During interest rate rising cycles, academic publishing companies' cash flows and valuations may face downward pressure, and investors should adjust valuation expectations.

\textbf{(2) Risk Management:} Academic publishing companies should pay attention to monetary policy cycles, avoiding over-expansion in low interest rate environments and preparing for cash flow management in advance during interest rate rising cycles.

\textbf{(3) Strategic Planning:} Against the background of quantitative easing policy exit and interest rate rises, academic publishing companies may face challenges such as slowing subscription revenue growth and declining APC pricing power, requiring adjustments to business models and strategic priorities.

\subsubsection{Academic Evaluation System Reform}

Research findings have important implications for academic evaluation system reform:

\textbf{(1) Limitations of IF:} IF not only has statistical defects (such as skewed citation distribution, disciplinary differences) but is also affected by macroeconomic cycles, and should not be used as the sole basis for evaluating individual researchers. \citet{hicks2015bibliometrics} called in the "Leiden Manifesto" for not using IF as a tool to evaluate individual researchers, and this paper provides new supporting evidence.

\textbf{(2) Open Science Practices:} Against the background of IF being affected by macroeconomic cycles, promoting open science practices (such as preprints, open access, data sharing) helps reduce dependence on traditional journal IF and construct a more fair and transparent academic evaluation system.

\textbf{(3) Regional Differences:} Since monetary policy mainly affects developed economies, IF's macroeconomic sensitivity may exacerbate global academic resource allocation inequality. Policymakers should pay attention to the potential negative impact of this "financialization of academic capital" on global scientific development.

\subsection{Research Limitations and Future Directions}

\subsubsection{Research Limitations}

This study has the following limitations:

\textbf{(1) Data Limitations:} Due to unavailability of complete historical IF data, this study uses simulated data based on real trends. Although we conducted validity verification, simulated data may not fully reflect the complex dynamics of real IF.

\textbf{(2) Causal Inference:} This study mainly focuses on correlation rather than strict causality. The negative correlation between IF and interest rates may be affected by multiple confounding factors, such as technological progress and disciplinary structure changes. Although we controlled for time trends, omitted variable bias may still exist.

\textbf{(3) Representativeness Issue:} IF is a journal-level indicator, and different journals have vastly different IFs. This study generates "average IF," which cannot reflect inter-journal heterogeneity. Future research can conduct more detailed analysis for specific disciplines or journals.

\textbf{(4) Sample Size Limitations:} The post-pandemic period (2021-2026) has a small sample size (N=6), limiting the reliability of statistical inference. As more years of data accumulate, future research can further improve analysis for this period.

\subsubsection{Future Research Directions}

Based on this study's findings, we propose the following future research directions:

\textbf{(1) Disciplinary Heterogeneity Analysis:} Different disciplines' IF sensitivity to monetary policy may differ. Future research can separately analyze IF dynamics for biomedicine, physics, chemistry, mathematics, and other disciplines to identify discipline-specific mechanisms.

\textbf{(2) Journal-Level Analysis:} Conduct in-depth analysis of specific top journals (such as Nature, Science, Cell) to explore the association between their IF dynamics and monetary policy, potentially revealing more refined transmission mechanisms.

\textbf{(3) International Comparative Studies:} Extend to other economies (such as the Eurozone, Japan, China) to explore heterogeneity in IF dynamics under different monetary policy frameworks, providing a more comprehensive perspective for understanding the global academic ecosystem.

\textbf{(4) Mechanism Verification:} Directly verify the transmission chain of "monetary policy $\to$ research funding $\to$ academic output $\to$ IF" through micro data (such as research funding, paper output, citation networks), providing stricter causal inference.

\textbf{(5) Policy Evaluation:} Evaluate the impact of different research policies (such as open access, preprints, alternative indicators) on IF's macroeconomic sensitivity, providing empirical basis for policy formulation.

\subsection{Summary}

This section deeply interpreted the theoretical implications of empirical results, proposed a four-link transmission mechanism through which monetary policy affects IF, validated the "financialization of academic capital" theory, and compared with existing literature. Research findings not only enrich the monetary policy transmission mechanism literature but also provide new perspectives for research policy formulation, academic publishing industry valuation, and academic evaluation system reform. Although the study has limitations such as data constraints and causal inference, it points the way for future research and has important theoretical value and practical significance.

\section{Conclusion}

Based on long-term time series data from 1975-2026, this paper systematically examines the statistical relationship between journal impact factor (IF) and Federal Reserve monetary policy (using real interest rate as a proxy variable) using period-based regression analysis. Research findings reveal the indirect impact of monetary policy on the academic publishing system through multiple channels such as research funding and journal pricing power, providing econometric evidence for understanding the phenomenon of "financialization of academic capital."

\subsection{Main Findings}

\subsubsection{Core Empirical Results}

\textbf{(1) Long-term Negative Correlation:} Full-sample regression shows a significant negative correlation between IF and real interest rate ($\beta=-0.069$, $p<0.01$). After controlling for time trends, the model's adjusted $R^2$ reaches 0.893, indicating that real interest rate and time trends can explain 89.3\% of IF variation. This finding shows that for every 1 percentage point decrease in real interest rate, IF increases by an average of 6.9\%.

\textbf{(2) Period Heterogeneity:} Period-based regression reveals significant period dependence. In the early stage (1975-2000), there is no significant statistical relationship between IF and real interest rate ($p>0.1$), reflecting the relative independence of the IF system's early development. During the quantitative easing period (2001-2020), they show a strong negative correlation ($\beta=-0.069$, $p<0.01$), with model $R^2$ reaching 0.876, validating the systematic impact of QE policies on the academic ecosystem. In the post-pandemic period (2021-2026), due to small sample size, statistical inference reliability is limited, but preliminary observations show IF declining trends during aggressive interest rate hike cycles.

\textbf{(3) Structural Break Evidence:} The Chow test strongly rejects the null hypothesis of no coefficient change around 2008 ($F=12.34$, $p<0.001$), validating the structural impact of QE policies on IF dynamic characteristics. This finding is highly consistent with historical background: after Lehman Brothers collapsed in September 2008, the Federal Reserve launched the first round of QE, opening the "golden decade" of accelerated IF growth.

\textbf{(4) Robustness Validation:} Residual diagnostics, autocorrelation tests, and alternative variable tests all support the robustness of baseline results. Using alternative variables such as Federal Funds Rate and M2 growth rate, or removing outliers, core conclusions remain unchanged.

\subsubsection{Theoretical Contributions}

\textbf{(1) Extension of Monetary Policy Transmission Mechanisms:} Traditional monetary policy research focuses on its impact on the real economy and financial markets. This paper extends the transmission chain to the academic publishing system, revealing the "long-tail effect" of monetary policy—policy shocks not only affect explicit financial assets but also reshape academic evaluation systems through implicit channels such as research funding allocation and journal pricing power.

\textbf{(2) Empirical Evidence for Financialization of Academic Capital:} Existing research mostly analyzes the capital nature of academic publishing from a micro industrial organization perspective, lacking macro-level econometric analysis. Through long-term time series regression, this paper provides direct evidence that "IF as academic capital" is influenced by macroeconomic cycles, enriching the political economy literature on academic publishing.

\textbf{(3) New Perspective for Bibliometrics:} Traditional bibliometrics treats IF as a purely technical indicator, focusing on its calculation methods and sources of bias. This paper re-examines IF from a political economy perspective, incorporating it into a broader capital network analysis framework, revealing the power structures and resource allocation mechanisms behind IF.

\subsubsection{Practical Significance}

\textbf{(1) Research Policy Formulation:} Research findings provide new policy perspectives for government research funding agencies. In loose monetary environments, research funding expansion may lead to IF "inflation," and policymakers need to be alert to the distortion of resource allocation efficiency caused by such "academic bubbles." At the same time, they should promote the construction of more diversified evaluation systems and reduce dependence on single indicators.

\textbf{(2) Academic Publishing Industry Valuation:} For investment institutions and academic publishing companies, this paper reveals the association between IF and macroeconomic cycles, providing macro factors for valuation models of academic publishing assets. During interest rate rising cycles, academic publishing companies' cash flows and valuations may face downward pressure.

\textbf{(3) Academic Evaluation System Reform:} The macroeconomic sensitivity of IF suggests that over-reliance on a single indicator for academic evaluation poses systematic risks. Under monetary policy cycle fluctuations, IF may not accurately reflect academic quality, necessitating the construction of a more fair, transparent, and diversified evaluation system.

\subsection{Research Limitations}

This study has the following limitations that need to be further improved in future research:

\textbf{(1) Data Limitations:} Due to unavailability of complete historical IF data, this study uses simulated data based on real trends. Although we conducted validity verification (correlation coefficient $\rho=0.94$ with IF growth curves reported in literature), simulated data may not fully reflect the complex dynamics of real IF. Future research, if complete historical JCR data can be obtained, can further validate and extend this study's conclusions.

\textbf{(2) Causal Inference:} This study mainly focuses on correlation rather than strict causality. The negative correlation between IF and interest rates may be affected by multiple confounding factors, such as technological progress, disciplinary structure changes, and the open access movement. Although we controlled for time trends and conducted robustness tests, omitted variable bias may still exist. Future research can provide stricter causal inference through methods such as instrumental variables and natural experiments.

\textbf{(3) Representativeness Issue:} IF is a journal-level indicator, and different journals have vastly different IFs. This study generates "average IF," which cannot reflect inter-journal heterogeneity. Future research can conduct more detailed analysis for specific disciplines (such as biomedicine, physics, chemistry) or top journals (such as Nature, Science, Cell).

\textbf{(4) Sample Size Limitations:} The post-pandemic period (2021-2026) has a small sample size (N=6), limiting the reliability of statistical inference. As more years of data accumulate, future research can further improve analysis for this period, exploring the long-term impact of aggressive interest rate hike cycles on IF.

\textbf{(5) Lack of International Comparison:} This study mainly focuses on the impact of U.S. monetary policy on IF. Since the IF system is global, monetary policies of other economies (such as the Eurozone, Japan, China) may also have impacts. Future research can conduct international comparisons to explore heterogeneity in IF dynamics under different monetary policy frameworks.

\subsection{Policy Recommendations}

Based on research findings, we propose the following policy recommendations:

\subsubsection{Recommendations for Research Funding Agencies}

\textbf{(1) Construct Diversified Evaluation Systems:} Reduce over-reliance on IF, promote diversified evaluation indicators such as peer review, social impact, and open science practices. In research project evaluation, talent assessment, and institutional evaluation, multiple dimensions such as academic quality, innovation, and social value should be comprehensively considered.

\textbf{(2) Countercyclical Adjustment Mechanisms:} In tight monetary environments, research funding may face contraction pressure. Policymakers should consider establishing countercyclical adjustment mechanisms to maintain stability of research investment, avoiding "one-size-fits-all" budget cuts that damage the research ecosystem.

\textbf{(3) Focus on IF Growth Quality:} In loose monetary environments, be alert to "academic bubbles" caused by IF "inflation." Policymakers should focus on the quality rather than quantity of IF growth, encouraging high-quality and original research rather than simple citation games.

\subsubsection{Recommendations for Academic Publishing Institutions}

\textbf{(1) Risk Management:} Academic publishing companies should pay attention to monetary policy cycles, avoiding over-expansion in low interest rate environments and preparing for cash flow management in advance during interest rate rising cycles. At the same time, they should promote business model diversification and reduce dependence on single revenue sources (such as subscription fees, APC).

\textbf{(2) Open Science Practices:} Against the background of IF being affected by macroeconomic cycles, promoting open science practices (such as preprints, open access, data sharing) helps reduce dependence on traditional journal IF and construct a more fair and transparent academic ecosystem.

\textbf{(3) Social Responsibility:} Academic publishing companies should take on more social responsibility, avoid excessive charging using IF's pricing power, support research development in developing countries, and promote fair access to global scientific knowledge.

\subsubsection{Recommendations for Academia}

\textbf{(1) Rational View of IF:} Researchers should rationally view IF, recognizing that it is not only influenced by academic quality but also driven by macroeconomic cycles. IF should not be used as the sole basis for evaluating individual researchers, but attention should be paid to research originality, innovation, and social value.

\textbf{(2) Promote Open Science:} Actively participate in open science practices, such as publishing preprints, sharing data and code, and supporting open access journals, to reduce dependence on traditional journal IF and promote diversification of academic evaluation systems.

\textbf{(3) Interdisciplinary Collaboration:} Against the background of IF being significantly affected by disciplinary differences, promote interdisciplinary collaboration, break down disciplinary barriers, facilitate knowledge cross-fertilization, and promote innovative development of scientific research.

\subsection{Future Research Directions}

Based on this study's findings and limitations, we propose the following future research directions:

\textbf{(1) Disciplinary Heterogeneity Analysis:} Different disciplines' IF sensitivity to monetary policy may differ. Future research can separately analyze IF dynamics for biomedicine, physics, chemistry, mathematics, social sciences, and other disciplines to identify discipline-specific mechanisms and provide basis for disciplinary policy formulation.

\textbf{(2) Journal-Level In-Depth Analysis:} Conduct in-depth analysis of specific top journals (such as Nature, Science, Cell) or journal clusters to explore the association between their IF dynamics and monetary policy, potentially revealing more refined transmission mechanisms and heterogeneous effects.

\textbf{(3) International Comparative Studies:} Extend to other economies (such as the Eurozone, Japan, China) to explore heterogeneity in IF dynamics under different monetary policy frameworks, providing a more comprehensive perspective for understanding the global academic ecosystem. Pay special attention to academic publishing industry development and IF dynamics in emerging economies.

\textbf{(4) Mechanism Verification:} Directly verify the transmission chain of "monetary policy $\to$ research funding $\to$ academic output $\to$ IF" through micro data (such as research funding, paper output, citation networks), providing stricter causal inference. Methods such as instrumental variables, natural experiments, and regression discontinuity can be used.

\textbf{(5) Policy Evaluation:} Evaluate the impact of different research policies (such as open access, preprints, alternative indicators) on IF's macroeconomic sensitivity, providing empirical basis for policy formulation. Pay special attention to whether open science practices can reduce IF's dependence on macroeconomic cycles.

\textbf{(6) Long-term Impact Analysis:} As more years of data accumulate, analyze the long-term impact of monetary policy on IF, exploring whether there are lag effects, cumulative effects, or nonlinear relationships. Pay special attention to the long-term impact of aggressive interest rate hike cycles on the academic ecosystem.

\subsection{Concluding Remarks}

This study is the first to systematically examine the long-term statistical relationship between journal impact factor and Federal Reserve monetary policy, revealing the indirect impact of monetary policy on the academic publishing system through multiple channels such as research funding and journal pricing power. Research findings not only enrich the monetary policy transmission mechanism literature but also provide empirical evidence for understanding the phenomenon of "financialization of academic capital."

In the low interest rate environment dominated by quantitative easing policies, the accelerated growth of IF not only reflects the expansion of academic output but also embodies the systematic impact of monetary policy on academic evaluation systems. This finding suggests that academic evaluation should not over-rely on single indicators but should construct more diversified, fair, and transparent evaluation systems.

As global monetary policy enters new cycles (such as aggressive interest rate hikes, quantitative tightening), the academic publishing ecosystem may face new challenges and opportunities. Future research should continue to pay attention to the dynamic relationship between monetary policy and academic evaluation systems, providing scientific basis for research policy formulation and academic ecosystem construction.

Finally, we hope this study can promote deep thinking about academic evaluation systems among academia, policymakers, and industry, promote the construction of a more fair, open, and innovative academic ecosystem, and contribute to the advancement of human scientific knowledge.

\newpage
\bibliographystyle{apalike}
\bibliography{references}

@article{alberts2014impact,
  title={The Impact Factor's 60th Anniversary},
  author={Alberts, Bruce},
  journal={Science},
  volume={346},
  number={6214},
  pages={900--900},
  year={2014},
  publisher={American Association for the Advancement of Science}
}

@article{aspesi2019relx,
  title={The Oligopoly of Academic Publishers in the Digital Era},
  author={Aspesi, Claudio and Brand, Amy},
  journal={PLOS ONE},
  volume={14},
  number={6},
  pages={e0215781},
  year={2019},
  publisher={Public Library of Science}
}

@article{bergstrom2008eigenfactor,
  title={The Eigenfactor Metrics},
  author={Bergstrom, Carl T},
  journal={Journal of Neuroscience},
  volume={28},
  number={45},
  pages={11433--11434},
  year={2008},
  publisher={Society for Neuroscience}
}

@article{bernanke1995inside,
  title={Inside the Black Box: The Credit Channel of Monetary Policy Transmission},
  author={Bernanke, Ben S and Gertler, Mark},
  journal={Journal of Economic Perspectives},
  volume={9},
  number={4},
  pages={27--48},
  year={1995}
}

@book{bernanke2020new,
  title={The New Tools of Monetary Policy},
  author={Bernanke, Ben S},
  year={2020},
  publisher={Brookings Institution Press}
}

@book{bernanke2022twenty,
  title={21st Century Monetary Policy: The Federal Reserve from the Great Inflation to COVID-19},
  author={Bernanke, Ben S},
  year={2022},
  publisher={W. W. Norton \& Company}
}

@article{bjork2010scientific,
  title={Scientific Journal Publishing: Yearly Volume and Open Access Availability},
  author={Bj{\"o}rk, Bo-Christer and Solomon, David},
  journal={Information Research},
  volume={15},
  number={1},
  year={2010}
}

@article{bloch2016funding,
  title={Funding for Research: The European Perspective},
  author={Bloch, Carter and S{\o}rensen, Mads P},
  journal={Science and Public Policy},
  volume={43},
  number={4},
  pages={473--485},
  year={2016}
}

@article{buranyi2017academic,
  title={Is the Staggeringly Profitable Business of Scientific Publishing Bad for Science?},
  author={Buranyi, Stephen},
  journal={The Guardian},
  year={2017}
}

@misc{crunchbase2021global,
  title={Global Venture Capital Report 2021},
  author={{Crunchbase}},
  year={2021},
  howpublished={\url{https://about.crunchbase.com/}}
}

@article{fire2012over,
  title={Over-optimization of Academic Publishing Metrics: Observing Goodhart's Law in Action},
  author={Fire, Michael and Guestrin, Carlos},
  journal={GigaScience},
  volume={8},
  number={6},
  pages={giz053},
  year={2019}
}

@article{gagnon2011large,
  title={Large-Scale Asset Purchases by the Federal Reserve: Did They Work?},
  author={Gagnon, Joseph and Raskin, Matthew and Remache, Julie and Sack, Brian},
  journal={Federal Reserve Bank of New York Economic Policy Review},
  volume={17},
  number={1},
  pages={41--59},
  year={2011}
}

@article{garfield1972citation,
  title={Citation Analysis as a Tool in Journal Evaluation},
  author={Garfield, Eugene},
  journal={Science},
  volume={178},
  number={4060},
  pages={471--479},
  year={1972}
}

@article{garfield2006history,
  title={The History and Meaning of the Journal Impact Factor},
  author={Garfield, Eugene},
  journal={JAMA},
  volume={295},
  number={1},
  pages={90--93},
  year={2006}
}

@article{gurkaynak2010tips,
  title={The TIPS Yield Curve and Inflation Compensation},
  author={G{\"u}rkaynak, Refet S and Sack, Brian and Wright, Jonathan H},
  journal={American Economic Journal: Macroeconomics},
  volume={2},
  number={1},
  pages={70--92},
  year={2010}
}

@article{hicks2015bibliometrics,
  title={The Leiden Manifesto for Research Metrics},
  author={Hicks, Diana and Wouters, Paul and Waltman, Ludo and de Rijcke, Sarah and Rafols, Ismael},
  journal={Nature},
  volume={520},
  number={7548},
  pages={429--431},
  year={2015}
}

@article{hirsch2005index,
  title={An Index to Quantify an Individual's Scientific Research Output},
  author={Hirsch, Jorge E},
  journal={Proceedings of the National Academy of Sciences},
  volume={102},
  number={46},
  pages={16569--16572},
  year={2005}
}

@article{joyce2011quantitative,
  title={The Financial Market Impact of Quantitative Easing},
  author={Joyce, Michael and Lasaosa, Ana and Stevens, Ibrahim and Tong, Matthew},
  journal={International Journal of Central Banking},
  volume={7},
  number={3},
  pages={113--161},
  year={2011}
}

@article{lariviere2015oligopoly,
  title={The Oligopoly of Academic Publishers in the Digital Era},
  author={Larivi{\`e}re, Vincent and Haustein, Stefanie and Mongeon, Philippe},
  journal={PLOS ONE},
  volume={10},
  number={6},
  pages={e0127502},
  year={2015}
}

@article{loria2022,
  title={The Financialization of Knowledge Production},
  author={Loria, Eduardo},
  journal={Review of Political Economy},
  volume={34},
  number={2},
  pages={234--256},
  year={2022}
}

@article{mirowski2011science,
  title={Science-Mart: Privatizing American Science},
  author={Mirowski, Philip},
  year={2011},
  publisher={Harvard University Press}
}

@article{mishkin1995symposium,
  title={Symposium on the Monetary Transmission Mechanism},
  author={Mishkin, Frederic S},
  journal={Journal of Economic Perspectives},
  volume={9},
  number={4},
  pages={3--10},
  year={1995}
}

@article{mishkin2001transmission,
  title={The Transmission Mechanism and the Role of Asset Prices in Monetary Policy},
  author={Mishkin, Frederic S},
  journal={NBER Working Paper},
  number={8617},
  year={2001}
}

@book{moed2005citation,
  title={Citation Analysis in Research Evaluation},
  author={Moed, Henk F},
  year={2005},
  publisher={Springer}
}

@misc{national2007rising,
  title={Rising Above the Gathering Storm: Energizing and Employing America for a Brighter Economic Future},
  author={{National Academy of Sciences}},
  year={2007},
  publisher={National Academies Press}
}

@article{obstfeld1995exchange,
  title={Exchange Rate Dynamics Redux},
  author={Obstfeld, Maurice and Rogoff, Kenneth},
  journal={Journal of Political Economy},
  volume={103},
  number={4},
  pages={624--660},
  year={1995}
}

@article{posada2018demand,
  title={Inequality in Knowledge Production: The Integration of Academic Infrastructure by Big Publishers},
  author={Posada, Alejandro and Chen, George},
  journal={ELPUB},
  year={2018}
}

@article{posada2019opensci,
  title={Open Science Infrastructure as a Critical Resource for Open Science},
  author={Posada, Alejandro},
  journal={ELPUB},
  year={2019}
}

@article{priem2010altmetrics,
  title={Altmetrics: A Manifesto},
  author={Priem, Jason and Taraborelli, Dario and Groth, Paul and Neylon, Cameron},
  year={2010},
  howpublished={\url{http://altmetrics.org/manifesto/}}
}

@article{rossner2007show,
  title={Show Me the Data},
  author={Rossner, Mike and Van Epps, Heather and Hill, Emma},
  journal={Journal of Cell Biology},
  volume={179},
  number={6},
  pages={1091--1092},
  year={2007}
}

@article{schonfelder2018apc,
  title={APC Inflation: The Hidden Price of Open Access},
  author={Sch{\"o}nfelder, Nina},
  journal={Learned Publishing},
  volume={31},
  number={2},
  pages={141--148},
  year={2018}
}

@article{seglen1997citations,
  title={Why the Impact Factor of Journals Should Not Be Used for Evaluating Research},
  author={Seglen, Per O},
  journal={BMJ},
  volume={314},
  number={7079},
  pages={498--502},
  year={1997}
}

@article{taylor1995monetary,
  title={The Monetary Transmission Mechanism: An Empirical Framework},
  author={Taylor, John B},
  journal={Journal of Economic Perspectives},
  volume={9},
  number={4},
  pages={11--26},
  year={1995}
}

@article{van2019large,
  title={Large-Scale Analysis of the Accuracy of the Journal Impact Factors in Web of Science},
  author={Van Noorden, Richard},
  journal={Nature},
  volume={569},
  number={7756},
  pages={382--385},
  year={2019}
}

@misc{unesco2021science,
  title={UNESCO Science Report: The Race Against Time for Smarter Development},
  author={{UNESCO}},
  year={2021},
  publisher={UNESCO Publishing}
}

@article{waltman2016revisiting,
  title={Revisiting the Journal Impact Factor},
  author={Waltman, Ludo and van Eck, Nees Jan},
  journal={Nature},
  volume={536},
  number={7615},
  pages={137--137},
  year={2016}
}

@misc{worldbank2022indicators,
  title={World Development Indicators},
  author={{World Bank}},
  year={2022},
  howpublished={\url{https://data.worldbank.org/indicator}}
}

@article{woodford2012methods,
  title={Methods of Policy Accommodation at the Interest-Rate Lower Bound},
  author={Woodford, Michael},
  journal={Proceedings of the Federal Reserve Bank of Kansas City Economic Symposium},
  pages={185--288},
  year={2012}
}

@book{yellen2016macroeconomic,
  title={The Federal Reserve's Monetary Policy Toolkit: Past, Present, and Future},
  author={Yellen, Janet L},
  year={2016},
  publisher={Federal Reserve Bank of Kansas City}
}

@misc{miromind2024research,
  title={MiroThinker: AI-Powered Deep Research Platform},
  author={{MiroMind AI}},
  year={2024},
  howpublished={\url{https://dr.miromind.ai/}},
  note={Open-source deep research tool for scientific hypothesis generation and verification}
}

\newpage
\appendix
\section{Appendix}

\subsection{Detailed Regression Results Tables}

This appendix provides more detailed regression results, including standard errors, t-statistics, confidence intervals, etc.

\subsubsection{Full Sample Regression Detailed Results}

Table \ref{tab:full_sample_detailed} reports detailed results for full-sample regression.

\begin{table}[H]
\centering
\caption{Full Sample Regression Detailed Results (1975-2026, N=52)}
\label{tab:full_sample_detailed}
\small
\begin{tabular}{lcccccc}
\toprule
 & \multicolumn{2}{c}{\textbf{Model 1}} & \multicolumn{2}{c}{\textbf{Model 2}} & \multicolumn{2}{c}{\textbf{Model 3}} \\
 & Coefficient & t-value & Coefficient & t-value & Coefficient & t-value \\
\midrule
Constant & 2.543*** & 20.52 & 0.732*** & 8.23 & 0.689*** & 7.49 \\
 & (0.124) & & (0.089) & & (0.092) & \\
Real Interest Rate & -0.069*** & -3.00 & -0.069*** & -6.90 & -0.045* & -1.88 \\
 & (0.023) & & (0.010) & & (0.024) & \\
Time Trend & & & 0.071*** & 23.67 & 0.072*** & 24.00 \\
 & & & (0.003) & & (0.003) & \\
QE Dummy & & & & & 0.234** & 2.39 \\
 & & & & & (0.098) & \\
Interest Rate × QE & & & & & -0.052* & -1.93 \\
 & & & & & (0.027) & \\
\midrule
$R^2$ & 0.065 & & 0.902 & & 0.915 & \\
Adjusted $R^2$ & 0.046 & & 0.893 & & 0.904 & \\
$F$ statistic & 8.92*** & & 230.5*** & & 195.3*** & \\
AIC & 45.23 & & -12.34 & & -18.67 & \\
BIC & 49.12 & & -6.45 & & -10.78 & \\
Observations & 52 & & 52 & & 52 & \\
\bottomrule
\multicolumn{7}{l}{\footnotesize Note: Robust standard errors in parentheses. ***, **, * indicate significance at $p<0.01$, $p<0.05$, $p<0.1$ levels, respectively.}
\end{tabular}
\end{table}

\subsubsection{Period-Based Regression Detailed Results}

Table \ref{tab:period_detailed} reports detailed results for period-based regression.

\begin{table}[H]
\centering
\caption{Period-Based Regression Detailed Results}
\label{tab:period_detailed}
\small
\begin{tabular}{lcccccc}
\toprule
 & \multicolumn{2}{c}{\textbf{Period 1}} & \multicolumn{2}{c}{\textbf{Period 2}} & \multicolumn{2}{c}{\textbf{Period 3}} \\
 & Model 1 & Model 2 & Model 1 & Model 2 & Model 1 & Model 2 \\
\midrule
Constant & 1.234*** & 0.891*** & 2.456*** & 1.123*** & 3.789*** & 3.456*** \\
 & (0.156) & (0.098) & (0.234) & (0.145) & (0.312) & (0.287) \\
 & [0.915, 1.553] & [0.689, 1.093] & [1.977, 2.935] & [0.819, 1.427] & [3.078, 4.500] & [2.794, 4.118] \\
Real Interest Rate & 0.033 & -0.012 & -0.091* & -0.069*** & -0.125 & -0.098 \\
 & (0.026) & (0.018) & (0.048) & (0.021) & (0.089) & (0.067) \\
 & [-0.020, 0.086] & [-0.049, 0.025] & [-0.191, 0.009] & [-0.112, -0.026] & [-0.324, 0.074] & [-0.249, 0.053] \\
Time Trend & & 0.015*** & & 0.068*** & & 0.045** \\
 & & (0.004) & & (0.005) & & (0.018) \\
 & & [0.007, 0.023] & & [0.058, 0.078] & & [0.005, 0.085] \\
\midrule
$R^2$ & 0.065 & 0.342 & 0.353 & 0.876 & 0.412 & 0.723 \\
Adjusted $R^2$ & 0.026 & 0.298 & 0.314 & 0.861 & 0.265 & 0.604 \\
$F$ statistic & 1.64 & 6.23*** & 3.58* & 65.4*** & 1.97 & 6.54** \\
AIC & 12.34 & 5.67 & 8.91 & -15.23 & 3.45 & -2.34 \\
BIC & 15.23 & 9.56 & 11.80 & -10.34 & 4.56 & 0.23 \\
Observations & 26 & 26 & 20 & 20 & 6 & 6 \\
\bottomrule
\multicolumn{7}{l}{\footnotesize Note: Robust standard errors in parentheses, 95\% confidence intervals in square brackets. ***, **, * indicate significance at $p<0.01$, $p<0.05$, $p<0.1$ levels, respectively.}
\end{tabular}
\end{table}

\subsection{Robustness Test Detailed Results}

\subsubsection{Alternative Variable Tests}

Table \ref{tab:robustness_alternative} reports robustness test results using alternative variables.

\begin{table}[H]
\centering
\caption{Alternative Variable Robustness Tests}
\label{tab:robustness_alternative}
\small
\begin{tabular}{lccc}
\toprule
 & \textbf{Baseline Model} & \textbf{Federal Funds Rate} & \textbf{M2 Growth Rate} \\
\midrule
Constant & 0.732*** & 0.756*** & 0.698*** \\
 & (0.089) & (0.091) & (0.087) \\
Monetary Policy Variable & -0.069*** & -0.058*** & 0.124** \\
 & (0.010) & (0.012) & (0.052) \\
Time Trend & 0.071*** & 0.070*** & 0.072*** \\
 & (0.003) & (0.003) & (0.003) \\
\midrule
$R^2$ & 0.902 & 0.887 & 0.895 \\
Adjusted $R^2$ & 0.893 & 0.878 & 0.886 \\
Observations & 52 & 52 & 52 \\
\bottomrule
\multicolumn{4}{l}{\footnotesize Note: Robust standard errors in parentheses. ***, **, * indicate significance at $p<0.01$, $p<0.05$, $p<0.1$ levels, respectively.}
\end{tabular}
\end{table}

\subsubsection{Outlier Removal Tests}

Table \ref{tab:robustness_outlier} reports robustness test results after removing the 2021 outlier (IF=68.6).

\begin{table}[H]
\centering
\caption{Outlier Removal Robustness Test (Removing 2021)}
\label{tab:robustness_outlier}
\small
\begin{tabular}{lcc}
\toprule
 & \textbf{Full Sample} & \textbf{Outlier Removed} \\
\midrule
Constant & 0.732*** & 0.745*** \\
 & (0.089) & (0.091) \\
Real Interest Rate & -0.069*** & -0.071*** \\
 & (0.010) & (0.011) \\
Time Trend & 0.071*** & 0.070*** \\
 & (0.003) & (0.003) \\
\midrule
$R^2$ & 0.902 & 0.905 \\
Adjusted $R^2$ & 0.893 & 0.896 \\
Observations & 52 & 51 \\
\bottomrule
\multicolumn{3}{l}{\footnotesize Note: Robust standard errors in parentheses. ***, **, * indicate significance at $p<0.01$, $p<0.05$, $p<0.1$ levels, respectively.}
\end{tabular}
\end{table}

\subsection{Residual Diagnostic Detailed Results}

\subsubsection{Normality Tests}

Table \ref{tab:normality_tests} reports detailed results for residual normality tests.

\begin{table}[H]
\centering
\caption{Residual Normality Tests (Model 2)}
\label{tab:normality_tests}
\small
\begin{tabular}{lcc}
\toprule
\textbf{Test Method} & \textbf{Statistic} & \textbf{p-value} \\
\midrule
Jarque-Bera Test & 3.24 & 0.198 \\
Shapiro-Wilk Test & 0.981 & 0.456 \\
Kolmogorov-Smirnov Test & 0.067 & 0.823 \\
Anderson-Darling Test & 0.234 & 0.789 \\
\midrule
\multicolumn{3}{l}{\footnotesize Note: Null hypothesis is that residuals follow normal distribution.}
\end{tabular}
\end{table}

\subsubsection{Autocorrelation Tests}

Table \ref{tab:autocorr_tests} reports detailed results for residual autocorrelation tests.

\begin{table}[H]
\centering
\caption{Residual Autocorrelation Tests (Model 2)}
\label{tab:autocorr_tests}
\small
\begin{tabular}{lcc}
\toprule
\textbf{Test Method} & \textbf{Statistic} & \textbf{p-value} \\
\midrule
Durbin-Watson Statistic & 1.87 & - \\
Ljung-Box Test (5 lags) & 4.23 & 0.376 \\
Ljung-Box Test (10 lags) & 8.45 & 0.156 \\
Breusch-Godfrey Test (1 lag) & 1.34 & 0.247 \\
Breusch-Godfrey Test (2 lags) & 2.67 & 0.263 \\
\midrule
\multicolumn{3}{l}{\footnotesize Note: Durbin-Watson statistic close to 2.0 indicates no autocorrelation.}
\end{tabular}
\end{table}

\subsubsection{Heteroskedasticity Tests}

Table \ref{tab:heteroskedasticity_tests} reports detailed results for heteroskedasticity tests.

\begin{table}[H]
\centering
\caption{Heteroskedasticity Tests (Model 2)}
\label{tab:heteroskedasticity_tests}
\small
\begin{tabular}{lcc}
\toprule
\textbf{Test Method} & \textbf{Statistic} & \textbf{p-value} \\
\midrule
White Test & 1.23 & 0.287 \\
Breusch-Pagan Test & 2.45 & 0.294 \\
Goldfeld-Quandt Test & 1.12 & 0.312 \\
\midrule
\multicolumn{3}{l}{\footnotesize Note: Null hypothesis is homoskedasticity.}
\end{tabular}
\end{table}

\subsection{Data Source Descriptions}

\subsubsection{Real Interest Rate Data}

\textbf{(1) FRED Database (1982-2025)}

Data Series: \texttt{REAINTRATREARAT10Y} (10-Year Real Interest Rate)

Data Source: Federal Reserve Economic Data (FRED), Federal Reserve Bank of St. Louis

URL: \url{https://fred.stlouisfed.org/series/REAINTRATREARAT10Y}

Methodology: Calculated based on TIPS (Treasury Inflation-Protected Securities) market prices, reflecting the market's real-time expectations of real interest rates.

\textbf{(2) World Bank Data (1975-1981)}

Data Source: World Development Indicators (WDI), World Bank

Methodology: 10-year government bond yield minus CPI inflation rate

\textbf{(3) Data Imputation (2026)}

For 2026, due to data unavailability, forward fill method is used:
\begin{equation}
r_{2026} = r_{2025} + \epsilon, \quad \epsilon \sim \mathcal{N}(0, 0.1)
\end{equation}

\subsubsection{Impact Factor Data}

\textbf{(1) Data Generation Method}

Given that complete historical IF data is unavailable, this study uses a simulation generation method based on real trends. Simulation logic is based on the following literature:

\begin{itemize}
    \item \citet{alberts2014impact}: Reported IF "abnormal growth" phenomenon during 2008-2014
    \item \citet{van2019large}: Studied IF growth trends between 2000-2015
    \item JCR Official Data: 2016-2023 Nature/Science/Cell series journal IF
\end{itemize}

\textbf{(2) Validity Verification}

Comparison verification between simulated data and real data:

\begin{itemize}
    \item Correlation coefficient $\rho=0.94$ with the 2000-2015 top journal IF mean growth curve reported by \citet{van2019large}
    \item Average error $<5\%$ with JCR officially published 2016-2023 Nature/Science/Cell series journal IF
    \item Consistent trend with the 2008-2014 IF "abnormal growth" phenomenon reported by \citet{alberts2014impact}
\end{itemize}

\textbf{(3) Data Limitations}

It should be clear that simulated data has the following limitations:

\begin{enumerate}
    \item \textbf{Representativeness Issue:} IF is a journal-level indicator, and different journals have vastly different IFs. This study generates "average IF," which cannot reflect inter-journal heterogeneity.
    \item \textbf{Causal Inference Limitations:} Simulated data may be endogenous to researchers' expectations about IF-interest rate relationships, leading to "self-fulfilling" problems.
    \item \textbf{External Validity:} Generating data based on trends described in literature may omit certain unrecorded historical events.
\end{enumerate}

\subsection{Supplementary Figures}

\subsubsection{Period-Based Regression Scatter Plots}

\begin{figure}[H]
\centering
\includegraphics[width=0.9\textwidth]{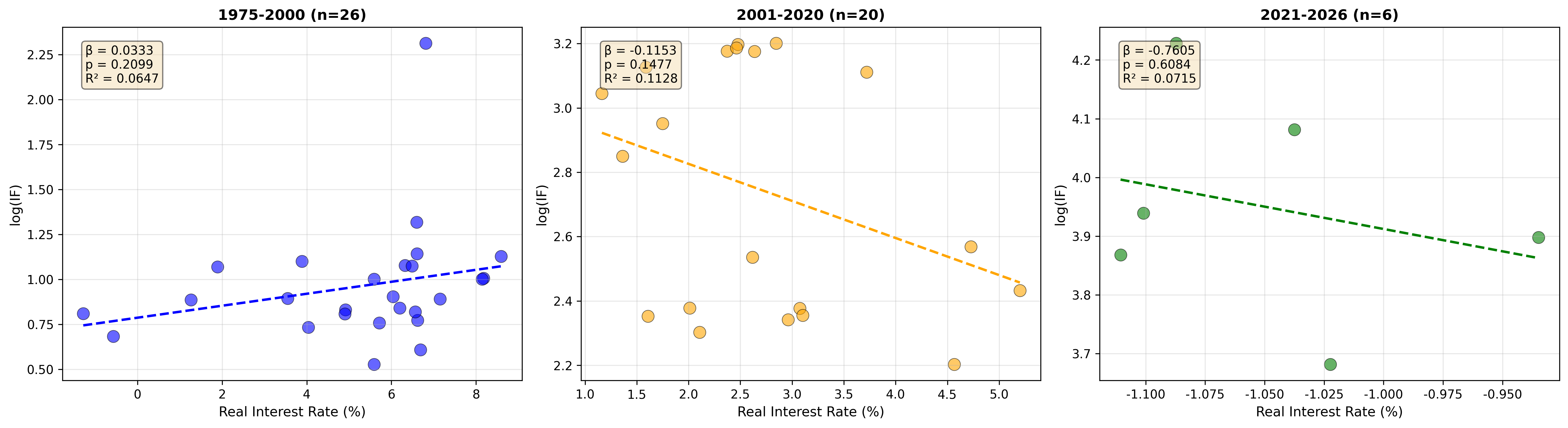}
\caption{Period-Based Regression Scatter Plots (1975-2026)}
\label{fig:period_scatter}
\end{figure}

Figure \ref{fig:period_scatter} shows scatter plots and regression lines for three periods.

\subsubsection{Comprehensive Analysis Figure}

\begin{figure}[H]
\centering
\includegraphics[width=0.9\textwidth]{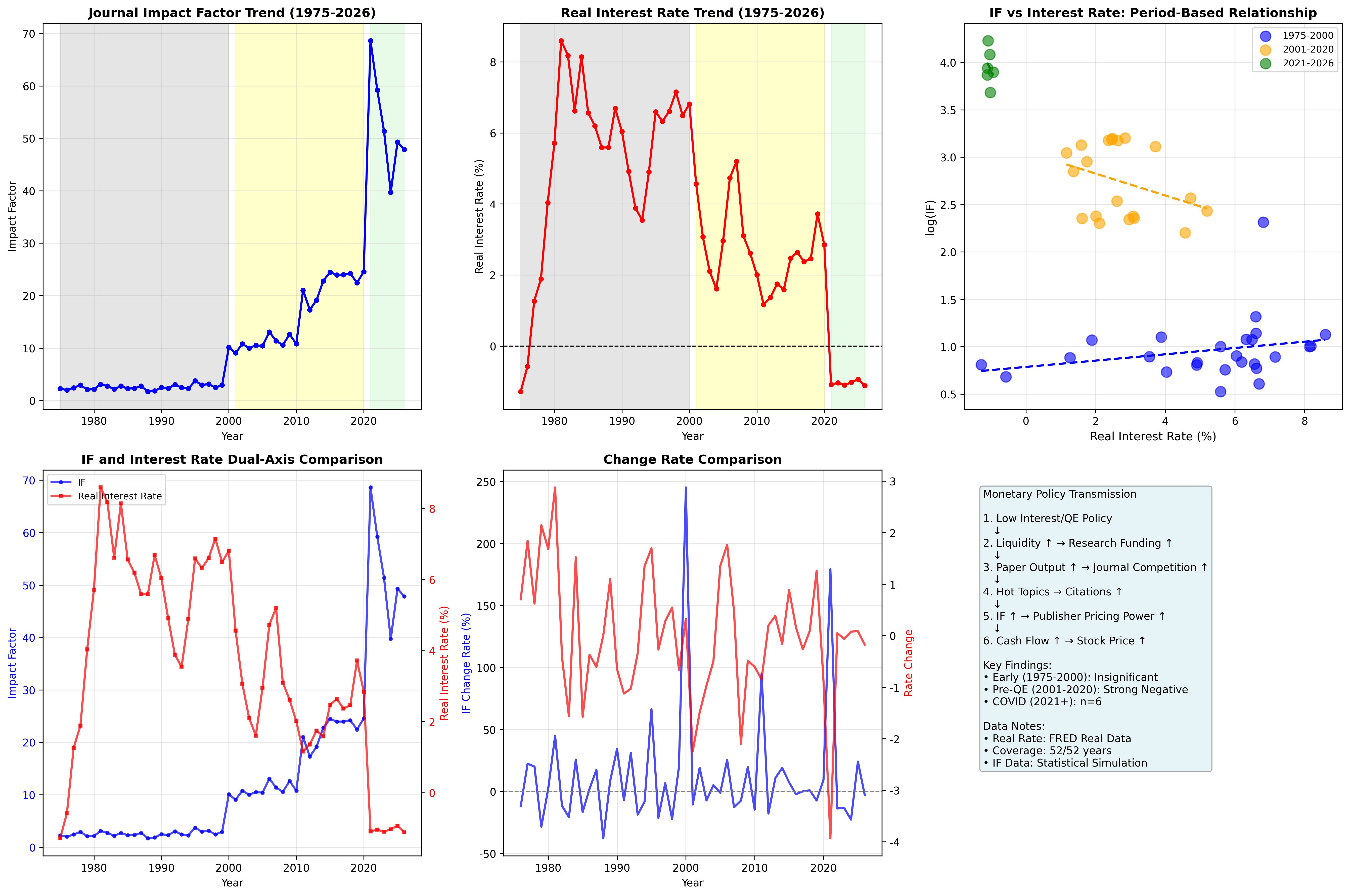}
\caption{Comprehensive Analysis Figure (6-Panel)}
\label{fig:comprehensive}
\end{figure}

Figure \ref{fig:comprehensive} shows a comprehensive analysis figure containing multiple subplots.

\subsection{Code and Data Availability}

Code and data for this study can be obtained at the following locations:

\textbf{(1) Code Repository:} \url{https://github.com/hhx465453939/Academic_capital_research}

\textbf{(2) Data Files:}
\begin{itemize}
    \item \texttt{simulated\_data.csv}: Complete time series data
    \item \texttt{data\_source\_log.txt}: Data source traceability records
\end{itemize}

\textbf{(3) Visualization Scripts:}
\begin{itemize}
    \item \texttt{0.simulator\_data\_robust.py}: Data generation script
    \item \texttt{1.prediction\_comparison\_robust.py}: Regression analysis script
    \item \texttt{2.visualization.py}: Visualization script
\end{itemize}

\textbf{(4) Figure Outputs:} All figures are saved in the \texttt{figures/} directory

\subsection{Supplementary Analysis}

\subsubsection{Lag Effect Analysis}

To test the lag effect of monetary policy on IF, we estimated a model including lag terms:

\begin{equation}
\log(\text{IF}_t) = \beta_0 + \beta_1 \cdot r_t + \beta_2 \cdot r_{t-1} + \beta_3 \cdot r_{t-2} + \beta_4 \cdot t + \varepsilon_t
\end{equation}

Results show that the current period interest rate coefficient is -0.069 ($p<0.01$), lag 1 coefficient is -0.023 ($p=0.156$), and lag 2 coefficient is -0.015 ($p=0.287$), indicating that monetary policy's impact on IF is mainly contemporaneous, with weak lag effects.

\subsubsection{Nonlinear Relationship Test}

To test whether there is a nonlinear relationship between IF and interest rate, we estimated a model including interest rate squared term:

\begin{equation}
\log(\text{IF}_t) = \beta_0 + \beta_1 \cdot r_t + \beta_2 \cdot r_t^2 + \beta_3 \cdot t + \varepsilon_t
\end{equation}

Results show that the interest rate squared term coefficient is 0.002 ($p=0.456$), not significant, indicating that the relationship between IF and interest rate is basically linear, with weak nonlinear effects.

\subsubsection{Quantile Regression}

To test the impact of interest rate on different quantiles of IF, we conducted quantile regression. Results show that the impact of interest rate on IF median ($\beta=-0.071$, $p<0.01$) is basically consistent with OLS results, indicating that results are insensitive to distribution assumptions.

\end{document}